%% file: fromQDTtoTD_Brownell2024.tex
\documentclass[%
 reprint,
 showkeys,
 amsmath,amssymb,
 aps,
 prb,
]{revtex4-2}

\usepackage{amsmath}% Include Am. Math. Soc. formatting package
\usepackage{amsfonts}% add more math fonts
\usepackage{amssymb}% add more math symbols
\usepackage{graphicx}% Include figure files
\usepackage{dcolumn}% Align table columns on decimal point
\usepackage{bm}% bold math
\usepackage{hyperref}% add hypertext capabilities

\begin{document}
\newcommand{\vect}[1]{\mathbf{#1}}% 
\newcommand{\fourvect}[1]{\bar{\vect{#1}}}% 

\title{On the transition from quantum decoherence to thermal dynamics \\ in natural conditions} 
\author{J. H. Brownell}
\email{JHBrownell@ThermodynamicsWithoutEntropy.org}
\noaffiliation
\date{September 10, 2024}

\begin{abstract}
A single mechanism, endemic to the standard model of physics, is proposed to explain wavefunction collapse, classical motion, dissipation, equilibration, and the transition from pure quantum mechanics through open system decoherence to the natural regime. Spontaneous events in the neighborhood of a particle disrupts correlation such that large many-particle states do not persist and each particle collapses to a stable mode of motion established by its neighbors. These events are the source of thermal fluctuation and drive diffusion. Consequently, evolution is not deterministic, unitary or classically conservative; diffusion toward a steady state occurs incessantly in every system of particles, though slowed under unnatural experimental conditions that suppress these events. Mean properties of a system evolve as particles jump between single-particle modes, producing observed transport laws and equilibrium properties without additional postulate or empirical factors. These modes are localized in dense material, yielding classical characteristics. Boltzmann's equal probability postulate is valid only when comparing results of nonrelativistic observers.
\end{abstract}

\maketitle

\section{Introduction}

Five observed processes appear at odds with quantum mechanics: decoherence, wavefunction collapse, classical motion, dissipation and equilibration. These anomalous phenomena are interrelated. Explanation of one affects how the others can be understood. One consistent theory is needed to resolve them all.

The standard model of physics (SMP) has developed over centuries of observation. Carefully controlled experiments reveal conservative fundamental physical forces as well as random spontaneous emission. Between spontaneous events within a well isolated system, such that spontaneous events among the surrounding particles do not substantially alter the potential felt by a system particle, internal microscopic motion is oscillatory, evolving by unitary time transformation, with physical variables represented by hermitian operators in a linear and deterministic Lagrangian equation of motion with time reversal symmetry. % (except for rare weak force events).

Absent spontaneous emission, an isolated system remains coherent. In this case, system state trajectories do not mix or diffuse, even in the limit of chaotic sensitivity, despite the appearance of random spatial distribution. An ensemble of identical systems would always be in the same state; every term in the ensemble density matrix maintains constant amplitude. Alternatively, if initial ensemble conditions are not uniform, then the amplitude of individual systems diverge, and the ensemble loses phase coherence, represented by decay of off-diagonal moment amplitudes in the ensemble density matrix. However, any initial state will eventually repeat, according to Poincare's recurrence theorem \cite{Poincare}. Ensemble coherence may decay for some period only to recover later. Even when this recurrence time is extremely long, half of a randomly prepared ensemble would exhibit growth in coherence while the other half exhibits decay. Furthermore, full phase coherence can be restored in principle, as demonstrated by spin and photon echoes \cite{Sechoes, Pechoes}.

Practical control is limited, though, and it may not be feasible to precisely reverse highly sensitive interaction, such as direct collisions. These cases appear effectively random and cause practically irreversible decoherence, such as diminishing strength of spin and photon echoes due to increasing rate of collisions.  Generally, random events, whether apparently or fundamentally stochastic as in spontaneous emission, diminish measured moments irreversibly. Only coherent interaction, such as with a laser beam, can produce a quantum state with a significant superposition of modes.

Quantum mechanics theory does not address why, in practice, every measurement produces one of a discrete set of possible values. The so-called measurement problem is that the system state apparently reduces, or collapses, during the detection process from a superposition to one of the eigenmodes of the operator representing the detector. Even though the system state may be known, measurement produces an uncertain result because wavefunctions occupy finite parameter space.

Neither does quantum mechanics explain why all systems are observed to equilibrate given enough time, losing coherence in the process. Dissipation, indicated by irreversible change in diagonal elements of the ensemble density matrix, thwarts oscillation, in glaring contradiction with conservative mechanics. The generally accepted reason for this universal tendency is that a non-mechanical property, entropy, always grows to a maximum in equilibrium, postulated in the Second Law of thermodynamics. Many mechanisms and methods have been explored to explain why dissipation occurs, such as classical dissipation functions \cite{Goldstein}, molecular chaos and coarse graining \cite{Uffink}, inherent dynamic probability of states \cite{Boltzmann, Gibbs}, apparent irreversibility \cite{Einstein}, subjective probability \cite{Jaynes}, many worlds \cite{MWI}, quantum Darwinism \cite{Zurek}, quantum chaos \cite{Srednicki1994, Alessio}, non-Hermitian operators \cite{Ashida}, and new stochastic interaction \cite{Bassi}. All of these proposals yield hints about what might be occurring. But each one relies variously on novel mechanisms outside of the SMP or effects assumed without consistent implementation or restricted to ideal gas and microsystems or on ensemble averaging, along with other assumptions. In each case, lack of derivation from SMP principles pushes the dissipation problem under different proverbial rugs.

Mark Srednicki concludes his paper on quantum chaos with key insight:
``More generally, in quantum mechanics, where time evolution is always
linear and therefore essentially trivial, the only place to encode the complexities of the
classical limit is in the energy eigenfunctions: that is where quantum chaos, like thermal
behavior, must be sought.'' \cite[p. 23]{Srednicki1994}.
I argue here that the source of quantum chaos has a more profound effect on the eigenstates themselves than considered to date, which opens an avenue to develop a complete theory of thermal dynamics.

The basic issue is that standard quantum mechanical analysis neglects fluctuation in the potential exerted by neighboring particles. This omission is tantamount to the principle of virtual work founding Lagrangian mechanics. This principle assumes that microscopic forces always balance so that they do zero net work on a particle over any time interval  \cite{Goldstein}. It is a necessary approximation for deriving Lagrangian equations of motion, which are conservative and time-symmetric as a result. Adding phenomenological dissipative terms, such as non-hermitian and stochastic operators in quantum mechanics or Rayleigh's dissipation function to the Lagrangian function, may suffice for estimating the trajectory of the microsystem or macroscopic object of interest, respectively. Such added terms are based on observed ``natural laws'' how the environment affects the system, without regard how the environment is affected in return. In which case, energy and momentum flow must be assumed to balance in some unspecified manner. These theories may serve a specific purpose but are incomplete.

After discussing current efforts, this paper proposes that spontaneous events dominate dynamics in complex systems under natural conditions and provide an endemic mechanism to explain the observed transition from quantum to classical regime as systems grow in complexity and density. A second paper shows how this mechanism causes diffusion and dissipative flow \cite{Brownell:transport}.

\section{Decoherence, reduction and eigenstate thermalization}

Practical ability to trap and control microsystems, having few active degrees of freedom, allows study of the transition from quantum to classical regimes by coupling a microsystem to another subsystem of greater complexity, i.e. more active degrees of freedom, usually called the environment in this context \cite{experiments}.
These experiments inspired open quantum system analysis to understand the repercussions of relaxing the pure quantum condition of isolation \cite{Isar, Vega, Ciccarello}. Two recent theories have garnered intense interest: Quantum decoherence theory (QDT) and the eigenstate thermalization hypothesis (ETH).

Quantum entanglement develops on interaction between any two arbitrarily defined subsystems. When an isolated system is two coupled trapped ions, the state of each ion oscillates as energy is transferred from one to the other and back without loss of coherence.  Greater complexity of either subsystem obscures this flow. From a coherent initial state, one may be said to cause decoherence in the other, due to entanglement, at a rate that increases with system complexity. Subsystem coherence appears to oscillate over the recurrence time period.

QDT describes evolution of a microsystem weakly coupled to its highly complex environment by unitary entanglement between system and environment \cite{Schlosshauer}. This scenario assumes that no spontaneous or effectively random events occur throughout the entire system, or more practically that an apparatus suppresses such events during an experiment. The coupling operator establishes an eigenbasis on the system.
Each eigen-component of the system state independently entangles with an environment state. Given zero initial correlation between system and environment, these components tend to diverge from each other in the vast parameter space of the environment, damping all interference terms in observable variable expectation values of the microsystem, after averaging over all possible states of the environment. This result is irreversible because dephasing occurs in the entangled components of the uncontrollable environment, unlike the dephasing mechanisms in spin echoes, for instance. Therefore, on average, the microsystem appears to evolve in a nonunitary manner to a classical ensemble of pure states over an interval, called the decoherence time, which shrinks with increasing complexity of the microsystem and environment and the coupling between them. In effect, the environment continually monitors the microsystem like a measurement device.

There are several issues with QDT:
(1) The entire system motion is unitary and oscillatory, implying that at some point coherence grows and the initial state eventually recurs. Decoherence occurs with certainty only if the initial state is maximally coherent. Any other initial state of partial coherence can progress to higher coherence or lower, contrary to evidence. This is the same issue faced by Boltzmann's H-Theorem \cite[Sec. 4.5]{Uffink}. Even if an ensemble distribution favors decoherence, some fraction of systems still would contradict experience.
(2) It is not evident how initial disentanglement can be prepared in the QDT model, unless it is also assumed that a thermalized microsystem state is unentangled. But this claim would contradict the QDT premise that thermal properties reflect rampant entanglement.
(3) QDT assumes that each subsystem is described by independent complete sets of eigenmodes so that averaging over environment states is unambiguous. Coupling would have to be so weak that interaction may be treated as a perturbation. This separation may be reasonable for microsystems but does not extend to more complex systems of interest.
(4) QDT presumes a projective measurement process, i.e. reduction or collapse, of environment states under observation. It is not evident why reduction should occur in one subsystem but not the other, particularly because bisecting a system is theoretically arbitrary in a comprehensive theory.
Therefore, QDT is practically limited to cases coupling a highly complex environment to a simple, low-energy microsystem that is not measured directly during an experiment, the cases for which it was developed.

The connection with macroscopic thermodynamics in isolated systems has been investigated most extensively through the ETH, which rests on the notion of quantum chaos: High energy eignstates of systems that exhibit classical chaos tend to be fairly equally separated, corresponding to divergent trajectories as a key feature of classical chaos \cite{Alessio}. Such spacing resembles the eigenstate distribution of random matrix ensembles \cite{Guhr}. By Berry's conjecture, an ensemble of systems exhibits a microcanonical distribution if the ensemble is randomly distributed over a set of energy eigenstates, defined as plane wave superpositions with a specific total energy \cite{Berry}. Observable ensemble expectation values then appear to thermalize due to dephasing of the off-diagonal elements of the observable operator and equal weighting of the eigenstates.

Even though dephasing is a unitary process in the ETH, the ensemble average changes irreversibly due to the random initial Berry condition. System processes are modeled as unitary dephasing intervals interspersed with random ``quenches'' caused by sudden shifts in one or more system parameters. This model matches the quasi-static ideal gas results in  Classical Thermodynamics and Statistical Mechanics. The standard thermodynamic relations follow from associating eigenstate changing processes with entropy change and eigenstate maintaining processes with external work.

While these results are encouraging, there are issues with ETH to resolve:
(1) Berry's conjecture describes an uncorrelated ``energy ensemble'' for each energy eigenvalue, which exhibits statistics of the microcanonical ensemble and is therefore fully thermalized. For example, this model matches the distribution of measured energy level spacing in thermalized nuclei \cite{Guhr}. In other words, the ETH is effectively circular by assuming \textit{a priori} that each energy eigenstate is a thermalized ensemble without explaining how unitary evolution produces an energy ensemble.
(2) Only the ensemble average exhibits irreversible thermal characteristics, contrary to observation of equilibration in every complex system. The typicality argument, that systems in a typical initial state evolves like the ensemble, doesn't suffice because, while most equilibrium states fall in a narrow range, the theory aims to understand the thermalization process of the atypical non-equilibrium states.
(3) A critical assumption in statistical mechanics theory, also employed in ETH, is equality between the long time average of unitary evolution of a single system and the average over a presumed energy ensemble. This may be reasonably close in equilibrium, even though ergodic theorem has not been proven generally, but there is no supporting evidence that this equality holds out of equilibrium.
(4) Quenches are non-unitary processes. No internal source of quenches is allowed in ETH, which is logical because a system cannot quench itself, otherwise motion could be perpetual. To be consistent, quenches must originate outside of the universe. They somehow disturb neither the perfect coherence of system eigenstates nor the perfect incoherence of the energy ensemble. No mechanism is identified. Like QDT, ETH is focused on practical microsystem analysis rather than comprehensive description of system and environment.
(5) Berry's conjecture is limited to the low density and high energy regime when particle wavelengths are small enough to react to chaos-inducing features in the potential profile and localized enough to move randomly.
It is not evident if it applies to dense or very low energy systems, the primary focus of current research.
(6) Each combination of quench and unitary dephasing process projects an ensemble onto a diagonal density matrix. Evolution by a sequence of projections produces rising entropy, and the standard thermodynamic relations, for the same reason that artificially imposed molecular chaos and coarse-graining breaks unitary time symmetry in Boltzmann's H-Theorem.
(7) Ultimately, ETH relies on state reduction to produce thermal behavior because a diagonal density matrix is interpreted as a statistical mixture of systems in a pure eigenstate, rather than dephasing of superposition states.

Both QDT and ETH then must explain reduction to be valid.
In order to resolve the measurement problem, a quantum equation of motion must include both nonlinear and stochastic interaction in order to cause reduction \cite{Bassi}. Current theories designed to induce reduction augment the Schr\"{o}dinger equation with new mechanisms to satisfy this condition. One model called continuous spontaneous localization (CSL) in Ref.~\cite{Bassi} mimics both quantum and classical regimes, for select values of model parameters, by assuming that a new interaction generates energy with finite Gaussian spatial profile and stochastically varying strength. (CSL is distinct from Anderson localization of multi-particle coherent states caused by interference from spatial fluctuation in media \cite{Anderson}.)
This interaction in the linear Schr\"{o}dinger equation causes the wavefunction norm to shift in time. The Born rule requires that the wavefunction should be renormalized to the ``physical'' probability, which produces a nonlinear ``physical'' Schr\"{o}dinger equation with Hamiltonian dependent on the current state. The CSL model prefers the position basis in order to assure that macroscopic objects do not spread noticeably or exist in two places simultaneously. This localization process is not in itself a measurement, though the sensing process in living tissue provides the conditions for reduction and our observation of classical behavior.

While the CSL model yields outcomes consistent with our experience for select parameter values, some unknown new mechanism drives the localizing stochastic process of every particle independently of surrounding material. Key questions are how such a ubiquitous interaction fits consistently with the SMP, and why such an interaction has not been identified yet? The simplest explanation would be that the mechanism is already present in the SMP.

\section{Natural conditions}\label{sec:natural}

Traditional arguments about how quantum decoherence and dissipation and the classical limit occur have generally assumed that system eigenstates always exist in complete form. This is reasonable if the system is confined by steady boundaries and coupling is weak enough to neglect modes spanning the environment as well. System modes are then independent of what happens outside. Entanglement appears as correlated amplitudes of independent sets of modes. Yet such constraint only allows limited conditional conclusions.

How do normal modes of motion form in the first place? Stable modes do not exist in empty space and consequently there is no \textit{a priori} preferred basis. Eigenmodes in principle exist only for eternally constant conditions and so are imaginary mathematical tools. Mode formation must be dynamic between these two cases. Conventional eigenmode expansion analysis neglects the transitory periods before modes are well established by interference. This neglect, in effect, precludes dissipation.

Consider the transition when a particle enters an empty reflective cavity through a narrow port. The particle's wavefunction expands immediately on entering the cavity as if in free space according to Schr\"{o}dinger's equation. Cavity walls do not affect the wavefunction until the leading wave front hits a boundary. Overlapping reflected wavelets that interfere constructively eventually form stable, persistent modes after many round trips of the wave fronts. Energy, momentum and other properties are conserved through interference. Any waves that interfere destructively cannot maintain finite amplitude within the cavity, indicating that the particle leaks out through the port. A combination of these cases is an entangled state until reduced to one or the other. If the entrance port is subsequently blocked, and absent absorption and fluctuations in the walls, the particle would persist in a superposition of stable modes that asymptotically approach  eigenfunctions satisfying the cavity boundary conditions.

Further consider the same situation with interior elements as well. There is no physical distinction between material within the cavity walls and any material that resides inside. Particles penetrate and reflect from material as a consequence of interference. The notion of ``cavity'' itself is a convenient construct to simplify the action of particles making up the cavity walls. Any material inside also causes diffraction by interference, which alters the stable modes. (Any absorption renders modes unstable and transitory.) When sufficiently dense, interior material effectively obscures the cavity walls. Stable modes then form by reflection from this nearer material, establishing the mode boundary conditions. More distant material affects mode structure only when any intervening dense material is stable and ordered enough such that distant reflection contributes significantly to overall interference, such as for conduction electrons in a crystal lattice.

This view accounts for all scale and type of ``cavity,'' from atomic structure to the free space limit. It describes the transition when a particle enters a new environment or its environment evolves. If this transition is slow compared to mode formation, then the particle state adiabatically follows as the modes adapt. If the transition is comparable to or faster than mode evolution, then the particle evolves into a superposition of the new stable modes that would persist if the potential were constant in time thereafter.

The new modes may not be mutually complete with the old ones. For example, when a cavity shrinks quickly, the portion of the particle state now outside cannot be represented by the new modes. In this case, there is a chance that the particle is bumped to an unexpected state, depending on how interference resolves the modes outside, and would not likely revert to the prior state if the cavity expands back again.

Neglect of spontaneous events in QDT and ETH becomes less accurate with more particles involved, both within the system and through coupling with its environment. Spontaneous emission and subsequent absorption, as well as sensitive collisions among particles, generate effectively random shifts in the local potential felt within the system.
These are extrinsic, not intrinsic, quantum jumps \cite{Plenio} with short but finite jump time governed by interference \cite{Shulman, delaPena}.
Each event disrupts the evolution of neighboring particles. Extended multi-particle states and mode superpositions are particularly susceptible to small shifts producing destructive interference. Only modes that maintain constructive interference despite such fluctuation are stable during a process, i.e. a mode is stable if a reduced particle reduces back to the same mode after environmental fluctuation and weak collisions.
Random disruption tends to destroy correlation between particles and reduce them to single stable mode states.

In terms of decoherence and reduction theories, each particle acts both as a microsystem and as the environment for all of its neighbors, effectively causing every particle to evolve in a non-unitary manner at all times.  However, assumptions in decoherence theory that may be reasonable for well isolated microsystems fail with strong coupling, namely the independence of microsystem modes from the environment, initially uncorrelated microsystem and environment states, and separate set of observables for microsystem and environment.
In principle, every coupled neighbor exerts a disrupting influence so that their modes are interdependent; the state of a particle depends on that of its neighbors, which in turn depends in part on the state and history of that particle. The coupling range is typically tens to thousands or more particles deep, in line with the CSL range suggested in Ref.~\cite{Bassi} to mimic the classical regime. The rate of disruption is then thousands to billions times greater than the individual spontaneous and collision rates. Consequently, including spontaneous emission renders the particle equation of motion both stochastic an non-linear, as required for reduction. Rapid fluctuations in the local potential function produced by random events among neighbors may represent a multitude of stochastic localization processes envisioned in CSL theory, with effective reduction rate magnified by the number of processes.

From the perspective of ETH, mode evolution may be the source of the assumed random ensemble. Stochastic reduction of individual particles causes the state of a system of particles to fluctuate. This activity may resemble quantum chaos by continually altering the system Hamiltonian randomly over a relatively small range, corresponding to the static ensemble presumed in random matrix theory.
The ``energy eigenstate'' conjectured by Berry mimics the ensemble characteristics of this fluctuation in a closed system (with constant energy) once it has settled to equilibrium. Open systems fluctuate in energy as well, which may be described as a random ensemble of energy eigenstates spread over energy.
However, the crucial distinction here is that all particles are uncorrelated. Representing their evolution as a stochastic ensemble of unitary system eigenstates may be complete mathematically but is unwarranted, because system eigenstates don't have time to resolve, and doing so obscures the key consequence that every system exhibits well-defined mean properties of its constituents.

A recent numerical study of eighteen coupled spins found that one particle obeys a thermal distribution if the state of each of the other particle interactions has Gaussian uncertainty \cite{Helbig}. This behavior is consistent with local mode evolution, yet is presented as justification for the ETH.  An explanation how such uncertainty develops in their unitary model is still needed to support the authors' claim to have derived statistical mechanics by quantum mechanics alone in non-integrable systems.

The QDT, CSL and ETH models support the view that emission, absorption and collision events occurring in the environment have a stochastic effect on a particle that disrupts its unitary evolution. Inversely, this view supports these models with an SMP mechanism.
There is a continuum of behavior from weakly coupled microsystems described by QDT and ETH to thermalization of strongly coupled subsystems in which every particle state reduces much faster than mode changing events occur and every particle exists in one currently stable mode practically all of the time.
Systems evolve predicably in this thermodynamic limit, even though the trajectories of individual particles fluctuate and mix, and so may seem microscopically chaotic. Classical chaos refers to conditions in which system macroscopic properties are highly sensitive to the initial mean state.

To summarize: Quantum modes of motion resolve by interference as particles traverse their available space, which is governed by surrounding particles. Mode resolution takes many round trips. Each particle state is highly sensitive to the configuration of its neighbors.  Any interaction causes phases shifts resulting in new interference patterns conserving energy and momentum. Eigenmodes of an entire complex system do not generally resolve before random particle motion irreversibly alters the potential landscape. Electron valence modes supported by a stable lattice of ions are a notable exception. Any group of fundamental particles that remain phase coherent for some time, like many-body scars in the ETH model \cite{Moudgalya, Chandran}, may be treated as composite particles for thermodynamic analysis. For example, nuclei, atoms, molecules, Cooper pairs, and many other quasiparticles are stable composite particles at progressively lower energy density threshold. Any interaction that repeatedly disrupts phase continuity, such as practically random mode and phase changing collisions as well as emission and absorption, produce interference that favors one mode by the CSL model. Only single particle modes are robust enough to be stable in natural conditions that quickly destroys any correlation and entanglement between particles.

This explanation suggests that reduction and localization of single particle states is endemic, strengthening with greater particle density and degrees of freedom. We observe macroscopic objects to be in one classical position because particle spacing is microscopic in condensed matter and the particles themselves contain their neighbors. The uncertainty in object position is the size of its particle modes, well below unaided human perception. Atoms and molecules are localized and stable relative to their center of mass for the same reason, and so act as composite particles. Fundamental quantum physics emerges only when reduction and diffusion are suppressed in all forms. The Shr\"{o}dinger cat paradox is resolved because superposition is a microscopic phenomenon; no natural, much less living, system can maintain a superposition of modes for any practical time span.

All complex physical systems equilibrate locally and globally because their constituents lose correlation through reduction and collisions.  The rate that uncorrelated particles transition from one mode to another is proportional to the population of the former. Therefore, over-populated modes, relative to a balanced state, tend to shed particles while under-populated modes gain. Far from equilibrium, each particle mode varies with local conditions and does not yet share characteristics that might be identified as a material phase. Mode structure evolves as the particle configuration shifts, and \textit{vice versa}, in a dance that gradually stabilizes toward steady mean mode population favoring a single material phase locally. This dance continues through externally driven shifts in local conditions.

The measurement problem may be understood as follows. A particle detector is a macroscopic device involving an enormous number of coupled yet uncorrelated particles. When a particle enters a detector, its state initially may resolve to a superposition of the detector entrance modes, yet is reduced to one of these modes by interactions with the material forming the detector. This reduction occurs very quickly, on the order of the detector material correlation time, destroying any entanglement such that only one mode survives to be recorded.

In this perspective, ``detection'' does not imply change in knowledge. Rather, reduction is a rapid, local quantum process that generates a readable macroscopic signal when occurring in an efficient detection device. Any device able to maintain unitary evolution cannot be thermal and would not be a detector in the traditional sense. Reduction is not specific to detectors and occurs continuously everywhere among collections of particles, whether inside or out of an arbitrarily designated system.

This conclusion impacts microsystem manipulation schemes, such as quantum computing \cite{QC}, in which a ``measurement'' process to initialize a definite state is usually assumed to be quick controlled step. The options are limited. Coherent interaction simply mixes the current state superposition of a qubit. Short and strong incoherent interaction reduces the state randomly to one of the eigenfunctions.  Spontaneous emission to a definite mode is apparently abrupt but not controllable in timing or phase. Lastly, thermal contact with the environment reduces each particle to its lowest available energy state, which would be uncorrelated particle modes as fluctuation disrupts formation of microsystem eigenstates.

The summary above also dispels any notion that there is a distinct class of classical phenomena or dynamics. The reduction mechanism originates with spontaneous emission, which is inherently quantum mechanical and integral to the SMP. Diffusion, dissipation, equilibration and thermal dynamics consequently follow in either direction of time \cite{Brownell:transport}.

Dynamic system models are tractable in either the QDT limit of a nearly isolated microsystem, or in the thermodynamic limit derived below. In the natural world described by thermodynamics theory, distinction of system from environment is an artificial convenience, coupling is rampant, boundaries are in continuous turmoil, and reduction is much faster than the time for ambient conditions to evolve and the interval between mode-changing events, i.e. strong collisions and absorption and emission of real particles. Complex systems practically evolve from one configuration of particle modes to another in quick jumps. As these mode changes represent transfer of energy, momentum and position, it is necessary to separate them from the reduction process in order to evaluate how a system evolves. The intermediate mesoscopic case, when the reduction rate is comparable to mode transition rates, does not simplify to concise equations of motion and may exhibit a combination of quantum and quasi-classical behavior.

\section{Measurement}\label{sec:measurement}

Practical measurement occurs over a finite time interval $\tau_{\text{meas}}$. There are four time scales to consider when conducting experiments: reduction time $\tau_{\text{red}}$, system correlation time $\tau_{\text{corr}}$, mode transition rates $1/\tau_{\text{mode}}$, and measurement time.
Outside of the experimental physics regime, $\tau_{\text{red}}$ is by far the smallest. In that case, ignore this fast reduction transition following mode changing events. Each particle effectively exists in one stable mode at any moment, not correlated with other particles, and relatively sudden transitions between modes.

Two further practical conditions are (1) measurement time must be short relative to significant change in state to capture system evolution but also (2) allow enough particles to transition among modes during a measurement for this number to be represented accurately by the average rate.

All system properties fluctuate with the system state. Thermodynamics theory focuses on the stable component, i.e. the mean values of measurements as the fluctuation tends to cancel out. Even though mean values still fluctuate for finite measurement time, the mean of repeated measurements has a well-defined asymptotic limit. There is a practical trade-off in choosing $\tau_{\text{meas}}$ between lower uncertainty in the measurement and less sensitivity to macroscopic change.
We naturally associate mean thermal properties with macroscopic, classical physics in the thermodynamic limit of large systems where the fluctuation amplitude is negligible.

Therefore, in the trajectory of any variable expectation value, identify the mean value averaged over a measurement period as the discernable macroscopic part and the difference as the thermal microscopic part. This distinction applies even to small and simple systems where fluctuation amplitude may be comparable or larger than the mean, in which case ``macroscopic'' trends still exist but may be difficult to discern.

A particle position expectation value $\vect{r}$ varies with each random transition to another mode. The microscopic part, $ \vect{r}_{\text{micro}}(t) = \vect{r}(t) - \vect{r}_{\text{macro}}$, is this fluctuating trajectory less the (macroscopic) mean position value $\vect{r}_{\text{macro}} = \langle \vect{r}(t) \rangle$, where angle brackets indicate averaging over the measurement period. Kinetic and potential energy terms expand into macroscopic, microscopic and mixed products. For example, the momentary kinetic energy of a single particle expands as $\text{KE} = \frac{1}{2}m \vert \vect{v} \vert^{2} = \frac{1}{2}m ( v_{\text{macro}}^{2} + 2\,\vect{v}_{\text{macro}} \cdot \vect{v}_{\text{micro}} + v_{\text{micro}}^{2} )$. The mixed terms contribute so long as the mean and fluctuation remain correlated. They average to zero for measurement times longer than the system correlation time ($\tau_{\text{corr}} < \tau_{\text{meas}}$).

Work done on a particle depends on how remote the source. Applied force fields generated by distant particles fluctuate randomly relative to a local particle trajectory, and therefore only the macroscopic term in $dW = F_{\text{applied}} \cdot d\vect{r}$ survives averaging. Fluctuation in the local configuration of neighboring particles, and the force they exert, is likely correlated to the test particle motion and some average microscopic work can shift its potential energy. Measured system energy is then comprised of two distinct components in natural conditions when correlation time tends to be very short:
\begin{equation}\label{systemenergy}
  \langle E_{\text{sys}} \rangle \approx \langle E_{\text{macro}} \rangle  + \langle E_{\text{micro}} \rangle \ .
\end{equation}

The system correlation time diminishes with increasing system complexity. Simple systems may not satisfy $\tau_{\text{corr}} < \tau_{\text{meas}}$, in which case this split becomes less distinct. They may be more accurately analyzed mechanically, employing Monte Carlo techniques to account for spontaneous events.

In addition to kinetic and potential energy, a stable particle carries energy associated with its formation. Fundamental particles carry rest mass energy and self field energy generated by its charge. A composite particle carries the formation energy of its constituent fundamental particles and their binding energy when brought together into its ground state configuration. Additionally, particles of a system settle into distinct and stable material phases in quasi-equilibrium. It is convenient for thermal process analysis to include in formation energy the mean binding energy, absent thermal motion, associated with each phase.  Formation energy is uniform and constant for all particles of a given constitution and material phase and can be computed theoretically. Constitution and phase identifies unique particle species and a particle's thermal properties.
(Binding energy varies smoothly near the interface between phases. Particles within this transition layer may be treated as the species of either adjacent bulk phase, with the difference in binding energy assigned to surface tension.)

Total system energy is the sum of constituent particle energy, which can be expressed in two ways when averaged for longer than the correlation time in quasi-equilibrium. Note that the total number $N_j$ of species $j$ in the system fluctuates as particles leave and enter and transform among species. This fluctuation is uncorrelated with each particle trajectory for large ($N_j \gg 1$), slowly varying systems, and mean energy due to fluctuating number averages to zero.

First method: Total kinetic energy $\text{KE} = \sum_j^{\text{Species}}\sum_k^{N_j}{\frac{1}{2}m_{j} \vert\vect{v}_{jk}\vert^2}$ with index $k$ specifying a particle.  The measured value $\langle \text{KE} \rangle$ splits into macroscopic $\sum_j^{\text{Species}} \langle N_j \rangle \frac{1}{2}m_j v_{\text{macro},j}^{2}$ and microscopic components, when the mean bulk flow velocity of species $j$ is defined as $\vect{v}_{\text{macro},j}=\sum_k^{\langle N_j \rangle} \langle \vect{v}_{jk} \rangle / \langle N_j \rangle$.  Similarly defining the mean bulk position for each species splits the total potential energy. Consequently, the total mean system energy splits into macroscopic and microscopic components. Microscopic energy represents the system formation and heat content $Q_{\text{sys}}$:
\begin{equation}\label{microenergy}
  \langle E_{\text{micro}} \rangle \approx \langle E_{\text{form}} \rangle  + \langle Q_{\text{sys}} \rangle
\end{equation}

Second method: The particle trajectory is evaluated as hopping among modes. Let $\tau_{jki}$ represent the duration that particle $k$ of species $j$ exists in mode $i$ during a measurement. The measured average energy of particle $k$ is then the sum of mode energies times the fraction of the measurement period $\tau_{\text{meas}}$ that it resided in each mode: $\langle E_{jk} \rangle = \sum_i^{\text{Modes}} \epsilon_{ji} \tau_{jki}/\tau_{\text{meas}}$, where $\epsilon_{ji}$ is the mean energy expectation value of a particle of species $j$ in mode $i$, including formation energy $\epsilon_{0j}$.  Now recognize $\sum_k^{\langle N_j \rangle} { \tau_{jki}/\tau_{\text{meas}} }$ as the mean mode occupation number $\langle N_{ji} \rangle$. Total mean system energy $\langle E_{\text{sys}} \rangle = \sum_j^{\text{Species}}\sum_k^{N_j} \langle E_{jk} \rangle$ becomes
\begin{equation}\label{systemenergy2}
\langle E_{\text{sys}} \rangle = \sum_j^{\text{Species}} \sum_i^{\text{Modes}} \epsilon_{ji} \langle N_{ji} \rangle \ .
\end{equation}
Each factor in this expression can be computed from first principles. Macroscopic energy can also be computed in the same manner by $\langle \vect{v}_{jk} \rangle = \sum_i^{\text{Modes}} \vect{v}_{ji} \tau_{jki}/\tau_{\text{meas}}$ and similarly for mean position. Likewise, flows of particles, energy and momentum are sums of mode occupation times quantum transition rate.

Given that $\langle E_{ji} \rangle$ is the absolute energy, the system heat content can then be determined by subtracting the macroscopic and formation energy components:
\begin{equation}\label{systemheat}
  \langle Q_{\text{sys}} \rangle \approx \langle E_{\text{sys}} \rangle - \langle E_{\text{macro}} \rangle  - \langle E_{\text{form}} \rangle
\end{equation}
Tracking heat content is the primary challenge in thermodynamic analysis. Friction (and dissipation generally) generated during a process can be computed directly from first principles by comparing change in heat and macroscopic energy, rather than accounted for indirectly through change in entropy. The next task is to estimate the mean mode energy and occupation in \eqref{systemenergy2}, which is feasible generally only in equilibrium.  A much larger set of parameters, some likely referring to prior states, is needed to describe a non-equilibrium state.

\section{Equilibrium distribution}

Practically, equilibrium is a state when the system no longer evolves on average, while recognizing that the instantaneous
behavior continues to fluctuate about this mean. Therefore, all mean flows are zero among modes, between species and between regions and the mean occupation of
each mode is steady in time. This macroscopically steady state
represents the set of all mode configurations that conform to these constraints, even as the system evolves among this set, and does
not imply existence in a pure system quantum eigenstate.

In principle, the mean distribution in equilibrium may be derived from any function that is highly sensitive to dynamically distinct particle configurations. Configurations that differ by dynamically irrelevant particle features do not alter flow and should be counted as degenerate. The number of possible dynamically distinct configurations $\Omega$ is a suitable and familiar function and is employed here. However, this measure is extensive and a condition for equilibrium should not favor a smaller system. Therefore, employ the intensive relative change condition $\left\langle\partial_t\Omega \big/ \Omega \right\rangle_t = \left\langle\partial_t \ln{\Omega}\right\rangle_t = 0$. The angle brackets represent time average over a measurement. The subscript is omitted hereafter.

All particles of the same constitution, energy and momentum respond to force in the same manner and so are dynamically equivalent. Therefore, group all particle modes by energy and momentum. Particles in different groups respond differently and processes exist that can distinguish them practically. A configuration $\{N_{ji}\}$, where $N_{ji}$ particles of a certain constitution, indexed by $j$, occupy the group of modes degenerate in energy and momentum indexed by $i$, is dynamically distinct from any other distribution among these mode groups. Index $j$ also represents unique species because regions tend to settle into uniform material phase.

This approach is similar to Boltzmann's combinatorial argument yet avoids objections identified in reviews such as Refs.~\cite{Uffink} and \cite{Frigg2021a}:
(1) It does not rely on Boltzmann's equal \textit{a priori} probability postulate or an implicit sampling of states to achieve equal probability. Both mode energy and momentum are necessary and sufficient to identify dynamically distinct particle states. Equal probability is deduced by relativistic invariance.
(2) It does not rely on the ergodic hypothesis and metric transitivity, which remain unproven  generally. %See Uffink "Foundations ..." section 4.4.2(1).
(3) It does not rely on any symmetry or notion of sufficient reason.
(4) It is valid for any particle interaction, and not limited to ideal gas. %See Uffink "Foundations ..." section 4.4.2(1).
(5) It does not involve phase space cell size. There is no modern consensus how to resolve this issue in Boltzmann's method. %See Uffink "Foundations ..." section 4.4.2(2).
(6) It is inherently dynamic, not a static maximal condition as stipulated by Boltzmann, and produces a quasi-equilibrium condition that indicates when the distribution practically approximates equilibrium. %See Uffink "Foundations ..." section 4.4.2(1).
(7) It does not require detailed balance in the Maxwell-Boltzmann sense that every path must balance.
(8) It applies to finite systems of many particles, as opposed to methods that resolve at the infinite size limit. ]
(9) It describes individual systems, and does not rely on ensemble statistics.

Internal flows among modes of the same species balance in equilibrium. Mean net flows to the outside also must be zero to achieve a steady state. Introduce total system particle number $N_j$, energy $E$ and momentum $\vect{P}$ by the trick of Lagrange multipliers, adding to $\ln{\Omega}$ terms that are identically zero: $\eta_j(N_j-\sum_{ji} N_{ji})$, $\beta_{\text{E}}(E-\sum_{ji} \epsilon_{ji} N_{ji})$, $\beta_{\text{p}x}(P_x-\sum_{ji} p_{xji} N_{ji})$, $\beta_{\text{p}y}(P_y-\sum_{ji} p_{yji} N_{ji})$ and $\beta_{\text{p}z}(P_z-\sum_{ji} p_{zji} N_{ji})$ with summation over mode groups.  Here $\eta_j$, $\beta_{\text{E}}$ and $\boldsymbol{\beta}_{\text{p}}$ are system-dependent functions that evolve along with mode energy spectrum while the system is out of equilibrium.
The equilibrium condition may then be expressed as
\begin{multline} \label{domegadt0}
0 \approx
\biggl\langle \partial_t \biggl\lbrack
\sum_j \eta_j \Bigl( N_j-\sum_i N_{ji} \Bigr)
+ \beta_{\text{E}} \Bigl( E-\sum_{ji} \epsilon_{ji} N_{ji} \Bigr) \\
+ \boldsymbol{\beta}_{\text{p}} \cdot \Bigl( \vect{P}-\sum_{ji} \vect{p}_{ji} N_{ji} \Bigr)
+ \ln{\Omega}
\biggr\rbrack \biggr\rangle \ .
\end{multline}

The total number of dynamically distinct configurations is the product of distinct arrangements in each group of degenerate modes. It's logarithm is then a sum of mode specific factors $\ln{\Omega}=\sum_{ji}\ln{\Omega_{ji}}$.
Neglecting the very slow change in Lagrange coefficients and mode spectrum ($\dot{\epsilon}_{ji}$ and $\dot{\vect{p}}_{ji}$) near equilibrium, \eqref{domegadt0} becomes
\begin{multline} \label{domegadt}
0 \approx
\sum_j \eta_j \langle \dot{N_j} \rangle
+ \beta_{\text{E}} \langle \dot{E} \rangle
+ \boldsymbol{\beta}_{\text{p}}\cdot \langle \dot{\vect{P}} \rangle \\
+ \sum_{ji} \left\langle \partial_t \ln{\Omega}_{ji} \right\rangle
- \left( \eta_j + \beta_{\text{E}} \epsilon_{ji} + \boldsymbol{\beta}_{\text{p}} \cdot \vect{p}_{ji} \right)
\langle {\dot{N}}_{ji} \rangle
\ .
\end{multline}
Rate of change is indicated by Newton’s dot accent $\dot{x} = \partial_t x$.

Individual mode population can only be stable if external particle and energy flows, and total force on the system, have no disrupting effect.  A quasi-equilibrium condition
may be stated as when the change in total particle number and energy and all applied forces are small enough, relative to the typical
fluctuation amplitudes in the modal terms, to produce a sufficiently
 stable population distribution for accurate analysis:
\begin{multline} \label{quasiEQcondition}
\biggl\vert
\sum_j \eta_j \langle \dot{N_j} \rangle
+ \beta_{\text{E}} \langle \dot{E} \rangle
+ \boldsymbol{\beta}_{\text{p}}\cdot \langle \dot{\vect{P}} \rangle
\biggr\vert
\ll \\
\biggl\vert
\sum_{ji} \left\langle \partial_t \ln{\Omega}_{ji} \right\rangle
- \left( \eta_j + \beta_{\text{E}} \epsilon_{ji} + \boldsymbol{\beta}_{\text{p}} \cdot \vect{p}_{ji} \right)
\langle {\dot{N_j}}_{i} \rangle
\biggr\vert
\ .
\end{multline}

These arguments apply to relativistic particles as well. Note that observers in any inertial reference frame observe the same set of modes, even if the mode profiles appear to be different. The number of particles existing in a mode is the same in every frame.  The summand in \eqref{domegadt} must then be invariant under Lorentz transformations. The number of configurations is invariant and so the other terms must be invariant as well. Parameter $\eta$ is invariant because it represents the difference in configuration number, as discussed below. If $\vect{p}$ is specified to represent canonical momentum, not mechanical momentum, then $\fourvect{p}=\{\epsilon,\vect{p}\}$ forms a relativistic 4-vector. Therefore, the four $\beta$ parameters must transform as a relativistic 4-covector $\fourvect{\beta}$ such that $\fourvect{\beta}_{\nu}\fourvect{p}^{\nu}_{ji} = \beta_{\text{E}}\epsilon_{ji} + \boldsymbol{\beta}_{\text{p}}\cdot\vect{p}_{ji}$ is an invariant 4-scalar. They also must be functions of the velocity $\vect{u}$ with which the observer moves relative to the frame in which the mode momentum $\fourvect{p}_i$ is evaluated. This condition is satisfied in flat spacetime when $\fourvect{\beta}=\beta_0\gamma_{u}\{1,-\vect{u}/c\}$ is proportional to the 4-velocity of the observer, given Lorentz factor $\gamma_{u}$.
For an observer analyzing a process in their own reference frame, $\vect{u}=0$ and $\fourvect{\beta} = \beta_0 \{1,0\}$. Only $\beta_{\text{E}}=\beta_0$ is not zero.  Two observers can compare their analyses directly if they move at nonrelativistic speeds ($\vert\vect{u}\vert \ll c$). However, four $\beta$ parameters are generally necessary to infer the distribution apparent to someone moving at relativistic speed.

First consider a system of one type of particle with a specific constitution. These results are later extended to describe mixed systems generally.

Fundamental particles are indistinguishable and obey specific constraints imposed by quantum phase symmetry.
Due to quantum interference, the Pauli exclusion principle requires that fermions never occupy the same mode as another
fermion of the same species. Therefore, for each group of degenerate modes, $N_{i} \leq g_{i}$ and $g_{i}!/N_{i}!\left( g_{i} - N_{i} \right)!$ is the number of unique
ways to arrange $N_{i}$ occupied modes plus $g_{i} - N_{i}$ empty
modes regardless of their order. The total number of fermion configurations is then
\begin{equation} \label{fermionconfigurations}
\Omega_{\text{fermion}} = \prod_{i = 1}^{\substack{\text{mode}\\\text{groups}}} \frac{g_{i}!}{N_{i}!\left( g_{i} - N_{i} \right)!} \ .
\end{equation}
On the other hand, bosons of the same species may exist together in the same mode. In this case,
\begin{equation} \label{bosonconfigurations}
\Omega_{\text{boson}} = \prod_{i = 1}^{\substack{\text{mode}\\\text{groups}}} \frac{\left( N_{i} + g_{i} - 1 \right)!}{N_{i}!\left( g_{i} - 1 \right)!} \ ,
\end{equation}
which counts the number of
unique ways to arrange $N_{i}$ particle interspersed with $g_{i} - 1$
dummy elements acting as mode dividers, regardless of order.

Both expressions converge to the same formula when $g_{i} \gg N_{i}$. Effectively
all particles become dynamically distinguishable when they rarely occupy
the same mode because there is no ambiguity as to which particle responds to interaction with a given mode. In other words, a lone particle is effectively
distinguishable by the mode it occupies. This condition defines the
classical regime. Quantum mechanical behavior becomes significant
when two indistinguishable particles are likely to occupy the same mode.

In our terrestrial experience, in the observer's rest frame, (a) only one $\beta$ parameter is pertinent, (b) total applied force on a system is not required to be zero in equilibrium because $\boldsymbol{\beta}_{\text{p}}=\vect{0}$, and (c) all modes degenerate in energy are then dynamically equivalent and equally populated. All of the modes with the same energy combine in \eqref{fermionconfigurations} and \eqref{bosonconfigurations} to produce the same formulas, but with index $i$ representing mode energy level with energy degeneracy $g_i$ regardless of momentum.

Note that both \eqref{fermionconfigurations} and \eqref{bosonconfigurations} reduce to $\Omega_{\text{fermion}} = \Omega_{\text{boson}} = 1$ when there is no degeneracy $(g_{i} = 1)$, in which case they are utterly insensitive to configuration and useless for our purpose. A relativistic perspective of a system of fundamental particles may have low or even no degeneracy.  Fortuitously, we can derive the rest frame distribution to high accuracy and then infer a relativistic perspective by the reasoning above.

In the following derivation of the equilibrium distribution it is assumed that $N_{i} \gg 1$ for every occupied mode energy group in order to employ Stirling's approximation for
$\ln{N!} = \sum_{n = 1}^{N} \ln n \approx \int_{1}^{N} dn \, \ln n \approx N\ln N - N$.
This assumption can be valid when $g_{i}$ is large even if the mean
occupation per mode is small because $N_i$ is the group occupation. Group degeneracy $g_{i}$ grows quickly with mode
energy in systems that allow motion in two and three dimensions. This
growth more than compensates for the tendency for particles to occupy
lower energy modes such that Stirling's formula is sufficiently
accurate, except possibly for the lowest level fermion modes which can accommodate few particles.

Consider the fermion case first. The boson case follows the same reasoning.
By Stirling's formula,
\begin{equation} \label{fermionoccupiedmodes}
\begin{split}
\ln\Omega_{\text{fermion}}
& = \sum_{i} \ln{g_{i}} - \ln{N_{i}} - \ln{((g_i - N_{i})!} \\
& \approx \sum_{i} g_{i}\left(\ln{g_{i}} - 1\right) - N_{i}\left(\ln{N_{i}} - 1\right) \\
& \quad \qquad - (g_i - N_{i})\left(\ln{(g_i - N_{i})} - 1\right) \ ,
\end{split}
\end{equation}
with sum over the spectrum of distinct mode energies.

Nonrelativisitic quasi-equilibrium condition \eqref{quasiEQcondition}, becomes
\begin{multline} \label{fermionquasiEQcondition}
0 \approx
\sum_{i} \left( \eta^{\text{FD}} + \beta \epsilon_{i} \right)\langle {\dot{N}}_{i} \rangle
+ \langle {\dot{N}}_{i}\ln{N_{i}} \rangle - \langle {\dot{N}}_{i}\ln{(g_i - N_{i})} \rangle
\ ,
\end{multline}
dropping the subscript from $\beta=\beta_{\text{E}}=\beta_0$. Again, change in $g_i$ is neglected because the mode spectrum varies slowly near equilibrium.

The rate of change in mode population is the aggregate of all inter-modal transitions. The two terms in the summand that are an average of the product of particle
numbers can be reduced by separating the mean from the rapid
fluctuation,
$N_{i} = \left\langle N_{i} \right\rangle + \delta_{i}$, so that
when the fluctuation is relatively small,
\begin{equation}
\begin{split}
\bigl\langle \dot{N_{i}}\ln N_{i} \bigr\rangle
& = \left\langle \dot{N_{i}} \ln{\Biggl\lbrack \left\langle N_{i} \right\rangle\left( 1 + \frac{\delta_{i}}{\left\langle N_{i} \right\rangle} \right) \Biggr\rbrack} \right\rangle \\
& \approx \bigl\langle \dot{N_{i}} \bigr\rangle \ln{\langle N_{i} \rangle} + \frac{\langle \dot{N_{i}}\delta_{i} \rangle}{\left\langle N_{i} \right\rangle} \ .
\end{split}
\end{equation}
The latter term on the right hand side is much smaller than the former
and quickly averages to zero because both the derivative and the
fluctuation amplitude are equally positive and negative over time and
are uncorrelated over the measurement period. Neglecting the latter term yields
\begin{equation} \label{FD_EQcondition}
0 \approx \sum_{i} \Bigl( \beta\epsilon_{i} + \eta^{\text{FD}} + \ln{\langle N_i \rangle} - \ln{\bigl(g_i - \langle N_i \rangle\bigr)} \Bigr) \, \langle {\dot{N}}_{i} \rangle \, .
\end{equation}

Despite the overall stability of total energy and particle number in
equilibrium, the individual particles continually change modes in an uncorrelated manner through
transitions and collisions, implying that
$\langle {\dot{N}}_{i} \rangle$ remain the largest terms in \eqref{quasiEQcondition}. Therefore, the
equilibrium condition is satisfied only if the factor in parentheses is
nearly zero for each group of degenerate modes:
\begin{equation}\label{FDequation}
\beta\epsilon_{i} + \eta^{\text{FD}} + \ln{\langle N_i \rangle} - \ln{(g_i - \langle N_i \rangle)} \approx 0 \quad \text{for all } i
\end{equation}
or
\begin{equation} \label{FDpop}
\left\langle N_{i}^{\text{FD}} \right\rangle \approx \frac{g_{i}}{e^{\beta\epsilon_{i} + \eta^{\text{FD}}} + 1} \ .
\end{equation}

Note that this result only depends on characteristics of a single mode because the rate $\langle {\dot{N}}_{i} \rangle$ factors completely in \eqref{FD_EQcondition}. The equilibrium distribution is robust in any system with enough particles, independent of the rate of particle interaction. Consequently, expressions relating to non-equilibrium flow cannot derive solely from equilibrium properties.

The equilibrium condition for a system of bosons similarly follows from
\eqref{bosonconfigurations}, given that Stirling's approximation requires
$N_{i} + g_{i} - 1 \gg 1$ so that
$N_{i} + g_{i} - 1 \approx N_{i} + g_{i}$, yielding the
Bose-Einstein formula:
\begin{equation}\label{BEequation}
\ln\left( \left\langle N_{i} \right\rangle + g_{i} \right) - \ln \left\langle N_{i} \right\rangle - \beta\epsilon_{i} - \eta^{\text{BE}} \approx 0 \quad \text{for all } i
\end{equation}
or
\begin{equation} \label{BEpop}
\left\langle N_{i}^{\text{BE}} \right\rangle \approx \frac{g_{i}}{e^{\beta\epsilon_{i} + \eta^{\text{BE}}} - 1} \ .
\end{equation}

In both cases, particles are dynamically indistinguishable among the degenerate modes and therefore evenly distributed among them.
Therefore, the mean population per mode is simply
$\left\langle N_{i} \right\rangle/g_{i}$.
Both Eqs.~\eqref{EQpopspecies} converge to the classical Maxwell-Boltzmann distribution at sufficiently high temperature, $\left\langle N_{i} \right\rangle/g_{i} \propto e^{-\beta\epsilon_{i}}$, when the mean number of particles occupying each mode is much less than one.

The above derivation of the equilibrium distribution assumes a system of
one species. Yet the only constraint on the set
of accessible modes is that they are stable in equilibrium, which is expected when the mean particle configuration and external conditions
are steady. These modes may include any interaction between particles
within the system and those outside. An exponential distribution must
result when the particle number and their energy are stable.
This result holds even when the set of particles under consideration
are embedded within a system containing other particle species, so long
as the net particle transformation rate and net energy flow between this
subset and the rest of the system is sufficiently low according to
\eqref{quasiEQcondition}.
(It is important to note that while particles may transform from one species into other species, they do not exchange subparticles. Any subparticles that may exist individually during a process are considered a distinct species.)
In other words, any subset of similar particles may be considered a system in
its own right. Particularly, this reasoning holds for every species in a
composite system. These subsets share the same temperature because the
number of configurations for a composite system is simply the product of
permutations for each species with a
common total energy constraint. Following the pure system derivation
closely, the mean quasi-equilibrium population per mode of each species with energy
$\epsilon_{ji}$ is
\begin{subequations} \label{EQpopspecies}
\begin{align}
\left\langle N_{ji}^{\text{FD}} \right\rangle \bigl/ g_{ji} & \approx \left( e^{\beta\epsilon_{ji} + \eta_j^{\text{FD}}} + 1 \right)^{-1} \ , \\
\left\langle N_{ji}^{\text{BE}} \right\rangle \bigl/ g_{ji} & \approx \left( e^{\beta\epsilon_{ji} + \eta_j^{\text{BE}}} - 1 \right)^{-1} \ ,
\end{align}
\end{subequations}
where the subscript index $j$ identifies the species of particle while
$i$ again refers to the mode.

Because any moving perspective counts the same number of particles in each mode, such an observer would tally the same average number over the same measurement interval, which is dilated relative to the rest frame. These formulas, therefore, may be extended to the relativistic regime by substituting $\beta\epsilon_{ji} \rightarrow \fourvect{\beta}_{\nu}\fourvect{p}^{\nu}_{ji} $.

Equations~\eqref{EQpopspecies} pertain all possible processes once an arbitrarily defined system has settled to quasi-stability by internal modal transition and by exchange of particles, energy and momentum with its environment.

\section{System statistics}

The remaining task is to determine the system constraint coefficients $\beta$ and $\{\eta_j\}$ in equilibrium in the rest frame.  The former sets an inverse energy scale, corresponding to mean particle energy per degree of freedom and expressed as temperature $T = 1/k_{\text{B}}\beta$ through the Boltzmann constant $k_{\text{B}}$. Temperature can be inferred from measured system properties or through contact with a calibrated thermometer.
The $\eta_j$ coefficients have no direct correspondence to our experience. Each $\eta_j$ parameter assures positive particle occupation in each mode with total number $N_j$. Some practical method is required to infer these coefficients from measurable system parameters.

A key feature of the distribution is the quantum-classical threshold near mode occupancy of one. Fermion occupation is suppressed by interference above roughly $\left\langle N_{j1} \right\rangle/g_{j1} = 1/2$, which defines a threshold temperature by $\beta\epsilon_{j1} = -\eta_{j}^{\text{FD}}$. Bosons, in contrast, tend to condense in the lowest energy modes at low temperature. Employing the same criterion, $\beta\epsilon_{j1} = \ln 2 - \eta_{j}^{\text{BE}}$ defines the boson threshold temperature.
It is tempting to express $\eta/\beta$ as an energy property (i.e. chemical potential) but doing so obscures the origin of $\eta$ and is inconsistent in this theory of absolute energy measure because $\eta$ varies from positive in the classical regime to negative in the quantum regime.

Equations~\eqref{EQpopspecies} provide the
means for expressing mean system thermodynamic properties as a
statistical average weighted by a relative probability distribution over
occupied states. Each mode group has well-defined physical properties. Therefore, each configuration of
particles has definite properties expressed through sensitivity of the
mode energy (and momentum in the relativistic regime) to change in measurable parameters characterizing the system.

This procedure can be shown easily in the classical regime and can be modified to produce quantum behavior as well.
Interpret $\left\langle N_{ji} \right\rangle/g_{ji}\left\langle N_{j} \right\rangle$
as the statistical weight of finding a particle of species $j$ in
the $i$th mode accessible to it, on average during a measurement in equilibrium. The
weight for each dynamically distinct state of particles
in equilibrium is then the product of the particle weights, $e^{- \beta E_{s}}$,
where the energy of the specific configuration
$\{ N_{ji} \}$, represented compactly by state index $s$, is
$E_{s} = \sum_{j}^{\text{species}}{\sum_{i}^{\text{modes}}{\epsilon_{ji} N_{ji} }}$.

The set of these weight values forms a distribution function over
particle configurations occupied during a measurement. This distribution normalized to one represents a probability for the system to be in a dynamically distinct state:
\begin{equation} \label{Pgen}
P_{s} = \frac{1}{Z} \, e^{- \beta E_{s}}.
\end{equation}
with normalizing system partition function defined as
\begin{equation} \label{partitionfunction}
Z( T,\{ \left\langle N_{j}\right\rangle \},\cdots ) \equiv \sum_{s}^{\text{states}}e^{- \beta E_{s}}.
\end{equation}
Here $T$ and $\{N_j\}$ are recognized explicitly as independent equilibrium system parameters. Other parameters such as volume and applied field amplitude may also be necessary to characterize a system sufficiently for thermodynamic analysis.

Define
\begin{multline} \label{etadef}
\eta_{j}( T,\{ \left\langle N_{k}\right\rangle \},V )
\equiv \ln{Z}( T,\{ \left\langle N_{1}\right\rangle,\ldots,\left\langle N_{j}\right\rangle,\ldots \},V ) \\
- \ln{Z}( T,\{ \left\langle N_{1}\right\rangle,\ldots,\left\langle N_{j}\right\rangle - 1,\ldots \},V ) \ ,
\end{multline}
which is generally dependent on the total particle number
$\{ N_k \}$ because the presence of interacting particles
affects the mode spectrum and therefore the normalization.
The volume is held constant in this difference to prevent external
work being done.
This difference mimics a function derivative for large $N_j$:
\begin{equation} \label{eta1}
\eta_{j} \approx \left. \frac{\partial\ln Z}{\partial \left\langle N_{j}\right\rangle} \right|_{T,V}.
\end{equation}

The partition
function may be considered the result of adding particles sequentially
to an empty volume,
\begin{equation}
Z( T,\{ \langle N_j \rangle \},V ) = \exp\Biggl\lbrack \sum_{\{ n_{j} \} = \{ 0 \}}^{\{ \langle N_j \rangle \}} \eta_{j}( T,\{ n_j \},V ) \Biggr\rbrack \ ,
\end{equation}
The function $e^{\eta_{j}}$ then may be interpreted as the
relative change in partition function due to the removal of one particle
of the $j$th species, and so is called the particle partition. The
property $\eta_{j}$ then acquires the awkward name of particle
partition logarithm (PPL). Note that $\eta_{j}$ is generally
dependent on all of the independent parameters and so cannot itself be
varied independently. For example, $\partial_{\eta_{j}}\ln Z$ is neither a
feasible operation nor is relevant analytically.

Consider a system of one species of fermions. Pauli exclusion
requires that only terms with single occupancy appear in the partition
function. Therefore, define the fermion partition function for $N$ constituents
as
\begin{equation} \label{fermionZ}
Z^{\text{fermion}}(T,N,V) \equiv \sum_{i_{1} \geq 1}^{\text{modes}}{\sum_{i_{2} > i_{1}}^{}{\cdots\sum_{i_{N} > i_{N - 1}}^{}e^{- \beta E_{s}}}} \ ,
\end{equation}
where $i_k$ indicates the mode of the $k$th particle. The sum index inequalities ensure that at most one fermion occupies a mode.

When the mode energies are independent of the configuration, which
describes ideally interacting particles, then
$E_{s} = \sum_{k = 1}^{N}\epsilon_{i_{k}}$ and the partition function
may be rearranged to isolate mode 1 terms as
\begin{multline} \label{fermionZseparated}
Z^{\substack{\text{ideal} \\ \text{fermion}}}
= \biggl(
e^{- \beta\epsilon_{1}}
\sum_{i_{2} > 1}
\cdots
\sum_{i_{N} > i_{N - 1}} e^{- \beta\sum_{k > 1}^{}\epsilon_{i_{k}}} \\
\quad + \sum_{i_{1} > 1}
\sum_{i_{2} > i_{1}}
\cdots
\sum_{i_{N} > i_{N - 1}} e^{- \beta \sum_{k > 1} \epsilon_{i_{k}}}
\biggr) \ .
\end{multline}
If $N_{1s}$ is the population of mode 1 in state $s$, the mean is $\langle N_{1}\rangle = \sum_s N_{1s}P_s$, which may be expanded as
\begin{equation}
\begin{split}
\left\langle N_{1}^{\substack{\text{ideal} \\ \text{fermion}}} \right\rangle
& = \frac{ e^{- \beta\epsilon_{1}}
\sum_{i_{2} > 1}
\cdots
\sum_{i_{N} > i_{N - 1}} e^{- \beta\sum_{k > 1} \epsilon_{i_{k}}}}{Z^{\text{ideal fermion}}} \\
& = \frac{1}{1 + \frac{\sum_{i_{1} > 1}^{}{\sum_{i_{2} > i_{1}}^{}{\cdots\sum_{i_{N} > i_{N - 1}}^{}e^{- \beta\sum_{k > 1}^{}\epsilon_{i_{k}}}}}}{e^{- \beta\epsilon_{1}}\sum_{i_{2} > 1}^{}{\cdots\sum_{i_{N} > i_{N - 1}}^{}e^{- \beta\sum_{k > 1}^{}\epsilon_{i_{k}}}}}} \ .
\end{split}
\end{equation}
Both nested sums in \eqref{fermionZseparated}
may be recognized as ``reduced'' partition functions, labeled
$Z'$, excluding mode 1 from the tally.  The latter sum involves $N$ particles,
\begin{equation} \label{reducedfermionZ}
{Z'}^{\substack{\text{ideal} \\ \text{fermion}}}(T,N,V)
\equiv \sum_{i_{1} > 1}^{} {\sum_{i_{2} > i_{1}}^{}{\cdots \sum_{i_{N} > i_{N - 1}}^{}e^{- \beta\sum_{k > 1}^{}\epsilon_{i_{k}}}}} \ ,
\end{equation}
while the former involves one less. Analogous to \eqref{etadef}, define the
reduced particle partition as
$e^{\eta'(T,N,V)} \equiv Z'(T,N,V)/Z'(T,V,N - 1)$ ,
so that
\begin{equation} \label{fermionmeanN1}
\left\langle N_{1}^{\text{ideal\ fermion}} \right\rangle
= 1 \Bigl/ \left( 1 + e^{\beta\epsilon_{1} + \eta{'}} \right) \ .
\end{equation}
This result applies to all modes because any mode may be labeled 1 in this derivation. It resembles the Fermi-Dirac distribution
\eqref{FDpop} but with reduced PPL
ensuring correct normalization. Pauli exclusion requires that many
energy levels are occupied even at low temperature so that this
difference between the reduced and full PPL value is insignificant under
all conditions in the thermodynamic limit. Therefore, $\eta'$ may be
replaced by $\eta$ to very high accuracy in
\eqref{fermionmeanN1} and further
equated with $\eta^{\text{FD}}$ in
\eqref{FDpop}:
\begin{equation} \label{fermionPPLrelation}
\eta^{\text{FD}} \approx \eta \ .
\end{equation}

For a system of $N$ boson particles of one species,
define the partition function as
\begin{equation} \label{bosonZ}
Z^{\text{boson}}(T,N,V) \equiv
\sum_{i_{1} \geq 1}^{\text{modes}}
\sum_{i_{2} \geq i_{1}}
\cdots
\sum_{i_{N} \geq i_{N - 1}} e^{- \beta E_{s}} \ ,
\end{equation}
where again $i_k$ indicates the mode of the $k$th particle. The sum index
inequalities allow multiple occupation and ensure that distinct
configurations are counted only once.

If the mode energies are
independent of the configuration, as for ideally interacting
particles, the partition function may be organized to isolate the
population of mode 1,
\begin{equation} \label{bosonZseparated}
\begin{split}
Z^{\substack{\text{ideal} \\ \text{boson}}}
& = \sum_{N_{1} = 0}^{N} e^{- N_{1}\beta\epsilon_{1}} \\
& \quad \times
\sum_{i_{N_{1} + 1} > 1}^{\text{modes}}
\sum_{i_{N_{1} + 2} \geq i_{N_{1} + 1}}
\cdots
\sum_{i_{N} \geq i_{N - 1}} e^{- \beta\sum_{k}^{}\epsilon_{i_{k}}} \\
& = \sum_{N_{1} = 0}^{N} e^{- N_{1}\beta\epsilon_{1}}{Z'}^{\substack{\text{ideal} \\ \text{boson}}}(T,V,N - N_{1}) \ ,
\end{split}
\end{equation}
where $Z'$ represents the reduced partition function omitting mode 1
analogous to \eqref{reducedfermionZ}.

The ideal boson PPL is independent of system particle number
outside of a narrow range just above threshold. Therefore, in most cases,
${Z'}^{\text{ideal boson}}(T,V,N - N_{1}) = e^{( N - N_{1} )\eta'(T,N,V)}$
and the mean population of mode 1 reduces to
\begin{equation} \label{indistbosonmeanN1}
\left\langle N_{1}^{\text{ideal boson}} \right\rangle = \frac{\sum_{N_{1} = 0}^{N}{N_{1}e^{- N_{1}\left( \beta\epsilon_{1} + \eta' \right)}}}{\sum_{N_{1} = 0}^{N}e^{- N_{1}\left( \beta\epsilon_{1} + \eta' \right)}} \ .
\end{equation}
Substituting $y = e^{- \beta\epsilon_{1} - \eta'}$ and identities
$\sum_{n = 0}^{N}y^{n} = ( 1 - y^{N + 1} ) / ( 1 - y )$
and $\sum_{n = 0}^{N}{n \, y^{n}} = y \, \partial_{y}\sum_{n = 0}^{N}y^{n}$
produces
\begin{equation} \label{BEmeanN1check}
\left\langle N_{1}^{\text{ideal boson}} \right\rangle = \frac{1}{1/y - 1} - \frac{N + 1}{ 1\big/ y^{N + 1} - 1}\ .
\end{equation}
This function transitions quickly between asymptotes at $\left\langle N_{1}^{\text{ideal boson}} \right\rangle=0$ and $N$ around $y=1$. The latter term is negligible for $y < 1 - 6/N$ or $\left\langle N_{1}^{\text{ideal boson}} \right\rangle < N/6$, which is true for all modes above the ground level.

The mean excited boson population resembles \eqref{BEpop} with reduced PPL. Again, the PPL may replace the reduced PPL to very high accuracy for excited modes.
Evaluation of formula \eqref{BEmeanN1check} for the ground mode is more subtle but the ground population must also match the Bose-Einstein value simply by subtracting the sum of all excited particles from $N$. Alternatively, an approximate form for all mode levels derived from Darwin-Fowler analysis of ideal bosons \cite{DF},
\begin{equation} \label{bosonPPLrelation}
\eta^{\text{BE}} \approx \eta + 1/N \ ,
\end{equation}
may be easier to work with. The $1/N$ term is significant only for the ground level near full saturation.

\section{Natural thermodynamics}

In natural conditions, mean mode population and mode transition rates determine all measured behavior, properties and flows. There are no concise formulas far from equilibrium. Near equilibrium thermodynamic analysis treats subsystems as rapidly thermalized while mode transitions between subsystems produce flows. All subsystem quasi-equilibrium thermodynamic properties may be computed as derivatives of the subsystem partition function with respect to the subsystem parameters.
Partition functions are computed by \eqref{fermionZ} and \eqref{bosonZ}. If not feasible theoretically, a partition function may be inferred from property measurements over a range of parameter values. The subsystem PPL and mean population distribution is then derived from this result by \eqref{etadef} and \eqref{EQpopspecies}.

It is important to recognize that while the partition function provides a concise way to relate equilibrium properties and to estimate fluctuation variance, it is not essential to thermodynamics and its association with ensemble probability has caused confusion.
State probability in this context means the relative number of dynamically similar configurations out of a set all possible equilibrium configurations. This is a static ratio, not the dynamic quantum probability per unit time to jump to another state that governs fluctuation statistics and flow. These two may be related, at least for ideal gas, through residence time only in the limit of infinite averaging period with the assumption of constant modes and ergodicity \cite{Uffink, Alessio, Deutsch}.
Out of equilibrium, modes shift and the likelihood that a configuration occurs depends on prior history. Residence time is then not normalizable as a probability function within a practical time interval that captures macroscopic motion. In other words, the notion of state probability is not a reliable tool for thermal dynamics. Yet Sec.~\ref{sec:measurement} implies that it is not necessary to assume that each particle traverses every mode or that all possible configurations occur over any time span. It is sufficient to recognize quantum mode transition rates as the only physical stochastic probability.
Also note that the quasi-equilibrium condition \eqref{quasiEQcondition} yields the same mean distribution whether the system is open or closed, adiabatically or fully isolated. These points together obviate any need to invoke microcanonical, canonical, grand-canonical, etc., stationary ensembles.

The current set of possible configurations does not represent an ensemble of individual systems. Rapid decorrelation among all particles breaks the notion that systems exist in a sequence of eigenstates and that the mean of a randomly prepared ensemble produce thermodynamic relations. If that were the case, a real system would wander over the entire range of configurations. Instead, the mean behavior of a given system remains relatively stable as all particles meander in opposing directions across many modes simultaneously. Ensembles and ergodicity are irrelevant concepts under natural conditions.

Internal and external fluctuation comes from the stochastic nature of quantum transitions, not summation over a random ensemble of unitary states (which still begs an explanation how the ensemble becomes random). It is reasonable that the breadth of a property distribution over equilibrium configurations is related to the fluctuation amplitude of that property. However, accurate computation of fluctuation and diffusion requires mode transition rates that are absent in the equilibrium distribution.

Detailed balance of mode transitions is expressed generally in three independent ways \cite{Brownell:transport}. These dynamic conditions supplant the static extremal conditions expressed in the Second Law. First, balance among transitions of particles of the same species within a subsystem, which is accounted for in \eqref{EQpopspecies}. Second, balance of particle transformations between subsystems of different species, accounted for in equilibrium kinetic conditions.
Third, balance between coupled, spatially adjacent systems implies that systems must be spatially uniform in equilibrium. Imbalance produces net spatial flow of particles, energy and momentum. Net flow equations match empirical ``natural laws'' when imbalance is slight.

The first balance condition was accessible to early pioneers in the field because the aggregate of mode transitions, represented solely in $\dot{N}_{ji}$, factors out of the equilibrium population equations \eqref{FDequation} and \eqref{BEequation}. Equilibrium is consequently very robust and characterized by steady-state parameters. But it also follows that equilibrium properties are not dynamic variables and cannot inform on the rate that a system can change. This issue is the reason why standard theory must postulate kinetic conditions, i.e. equal chemical potential, and relies on empirical transport equations for process analysis.

\section{Intermediate regime}

This section demonstrates how dissipation may appear in well isolated microsystems.
Trapped atoms are a common model for qubits in quantum computers.  A practical qubit must remain coherent far longer than the period needed to manipulate its state. Electromagnetic traps are continually improving how an atom can be maintained with long decoherence time. A common test induces Rabi oscillation between two stable modes of the atom with a coherent electromagnetic field \cite{Kuhr, Jones}. Probing one mode after various delay indicates an oscillating population. Dissipation causes the oscillation amplitude to degrade irreversibly over time, depending on the degree of isolation.

Weak thermal coupling with the qubit environment has two primary effects in this case. First, the atom settles to equilibrium when undisturbed. The initial state for a test is then the lower energy mode if the equilibrium temperature $k_{\text{B}}T$ is much less than the mode energy difference. Second, as discussed in Section~\ref{sec:natural}, quick mode changing events in the environment trigger reduction events in the atom stochastically, proceeding over several orbits of the atom modes. The latter mimic spontaneous emission, but with the average rate of reduction controlled by the degree of coupling.

The Rabi frequency $\Omega_{\text{R}}$ and mean reduction rate $\gamma$ are usually low compared to the mode frequency, in which case the transient reduction process can be approximated as instantaneous. The atom state may then be evaluated as follows.

An unperturbed two-level atom state evolves as $\vert \psi(t) \rangle = a(t) \vert 1 \rangle + b(t) e^{-i \omega_0 t} \vert 2 \rangle$, where $\hbar \omega_0$ is the energy difference between mode 1 and mode 2 and $\vert a \vert^2 + \vert b \vert^2 = 1$. A coherent field $E = A \cos (\omega t)$ excites dipole moment $p_{12}= \langle 1 \vert \hat{p} \vert 2 \rangle$ so that the Schr\"{o}dinger equation implies
\begin{align}
  \partial_t a &= -i b \Omega_{\text{R}}^{\ast} \left( e^{i (\omega - \omega_0) t} + e^{-i (\omega + \omega_0) t} \right) \\
  \partial_t b &= -i a \Omega_{\text{R}} \left( e^{i (\omega + \omega_0) t} + e^{-i (\omega - \omega_0) t} \right)
\end{align}
with Rabi frequency $\Omega_{\text{R}} = p_{12} \cdot A / 2\hbar$. An asterisk indicates complex conjugate. Neglect the fast oscillating terms which quickly average to zero with little effect. Only terms oscillating at the detuning frequency $\Delta = \omega - \omega_0$ remain. These linear coupled equations have solution $a(t) = f(t)$ and $b(t) = g(t)$ when $\{a(0)=1, b(0)=0\}$, or $a(t) = -g^{\ast}(t)$ and $b(t) = f^{\ast}(t)$ when $\{a(0)=0, b(0)=1\}$, defining
\begin{align}
  f(t) & = \left(\cos(\theta t) - i\frac{\Delta}{2\theta}\sin(\theta t) \right) e^{i \Delta t /2} , \\
  g(t) & = - i\frac{\Omega}{\theta}\sin(\theta t) e^{-i \Delta t /2} , \\
  \intertext{with}
  \theta & = \sqrt{\vert \Omega \vert^2 + \Delta^2/4} .
\end{align}
A strong pulse can send the atom quickly into a desired state.

For an initial state $\vert \psi(0)\rangle = \vert 2 \rangle$, the atom state would evolve to $\vert \psi(\tau)\rangle = -g^{\ast}(\tau)\vert 1 \rangle + f^{\ast}(\tau)\vert 2 \rangle$ after a time interval $\tau$ without reduction. For convenience, define the states $\vert\chi(t)\rangle= f(t)\vert 1 \rangle + g(t)\vert 2 \rangle, \vert\phi(t)\rangle = -g^{\ast}(t)\vert 1 \rangle + f^{\ast}(t)\vert 2 \rangle$ and  $F(t)=\vert f(t)\vert^2$ and $G(t)=\vert g(t)\vert^2$. For short intervals, $\gamma \tau \ll 1$ represents the mean probability of a reduction event during the interval. Now approximate the net effect as sequential unitary evolution and sudden reduction. This approximation affects the rate but not the character of exponential dissipation of the mode population.

There are three possible outcomes with different statistical weights for the atom state after the first interval when averaged over many repeated tests:
\begin{equation} \label{firstincrement}
\vert \psi(\tau)\rangle = \bigl\{(1-\gamma\tau)\phi(\tau), (\gamma\tau)G(\tau)\vert 1 \rangle, (\gamma\tau)F(\tau)\vert 2 \rangle \bigr\}
\end{equation}
Before a second reduction event, the atom evolves as
\begin{multline}
\vert \psi(t>\tau)\rangle = \bigl\{(1-\gamma\tau)\vert\phi(t)\rangle, \\
(\gamma\tau)G(\tau)\vert \chi(t-\tau)\rangle, (\gamma\tau)F(\tau)\vert \phi(t-\tau) \rangle \bigr\}
\end{multline}
There are nine possible outcomes after the second interval. Arranging these in order of statistical weight,
\begin{multline}
  \vert \psi(2\tau)\rangle = \Bigl\{
    (1-\gamma\tau)^2\phi(2\tau), \\
    (1-\gamma\tau)(\gamma\tau)\bigl\{ G(2\tau)\vert 1 \rangle, F(2\tau)\vert 2 \rangle, G(\tau)\vert \chi(\tau)\rangle, F(\tau)\vert \phi(\tau) \rangle \bigr\}, \\
    (\gamma\tau)^2\bigl\{ G^2(\tau)\vert 1 \rangle, G(\tau)F(\tau)\vert 2 \rangle, F(\tau)G(\tau)\vert 2 \rangle, F^2(\tau)\vert 1 \rangle \bigr\}
    \Bigr\}
\end{multline}
The number of outcomes triples with each interval yet follow a pattern. After $n$ intervals, the averaged state is
\begin{multline}
  \vert \psi(n\tau)\rangle = \biggl\{
    (1-\gamma\tau)^n\phi(n\tau), \\
    (1-\gamma\tau)^{n-1}(\gamma\tau)\bigl\{
      G(m\tau)\vert \chi((n-m)\tau)\rangle,  \\
      F(m\tau)\vert \phi((n-m)\tau) \rangle
      \bigr\}_{m=1}^{n}, \\
    (1-\gamma\tau)^{n-2}(\gamma\tau)^2\Bigl\{\bigl\{
      G(m\tau)G((k-m)\tau)\vert \phi((n-k)\tau) \rangle, \\
      G(m\tau)F((k-m)\tau)\vert \chi((n-k)\tau) \rangle, \\
      F(m\tau)G((k-m)\tau)\vert \chi((n-k)\tau) \rangle, \\
      F(m\tau)F((k-m)\tau)\vert \phi((n-k)\tau) \rangle
      \bigr\}_{k=m+1}^{n}\Bigr\}_{m=1}^{n-1},
    \cdots
    \biggr\}
\end{multline}

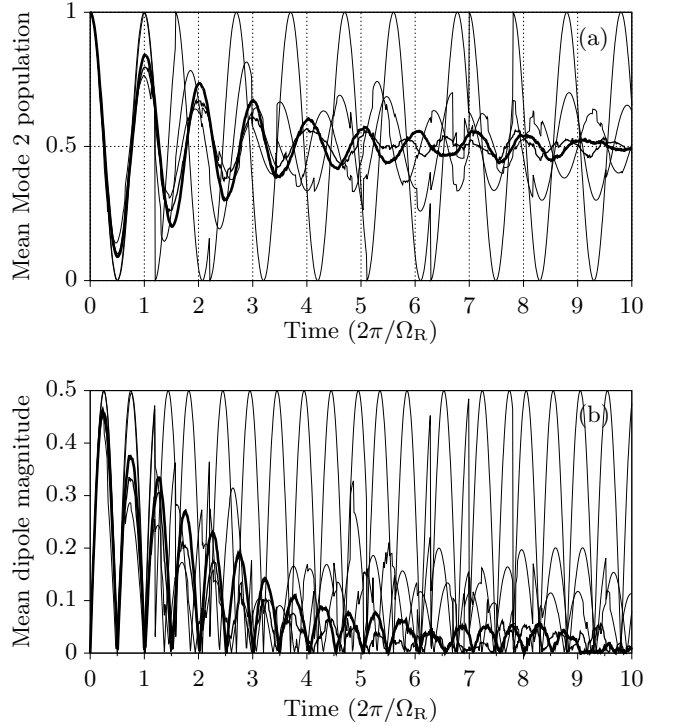
\begin{figure}
\centering
{\small
 \hspace*{-.5cm}
 \input{Fig_Level2pop.tex}
 \hspace*{-.5cm}
 \input{Fig_Dipole.tex}
}
\vspace{-6mm}
\label{fig:rabi}
\caption{Average mode 2 population and dipole magnitude of a 2-level atom driven by a resonant sinusoidal field coupling the two stable modes. The number of runs averaged increases with line thickness: 1, 4, 16, 64, 256. Fast reduction events occur at a mean rate of $0.25 \Omega_{\text{R}}$. Spontaneous emission and all other forms of dephasing are omitted. Note how the average of many runs appears synchronous with the ideal Rabi oscillation even though each run is not.}
\end{figure}

After many short intervals ($n \gg 1$ and $\gamma\tau \ll 1$) the leading coefficient approximates an exponential function, $(1-\gamma\tau)^n \approx e^{-\gamma n\tau}$. Each correction term is of order $\gamma\tau/(1-\gamma\tau)$ smaller than the previous one. With perfect thermal isolation of the atom from its environment ($\gamma = 0$), the leading term would yield an ideal constant amplitude Rabi oscillation of the mode population. When not isolated, reduction is extremely frequent in natural conditions, in which case the atom is critically damped, continually collapsing to the current state by the quantum Zeno effect \cite{Zeno} most of the time. Similar evaluation and conclusion applies generally to all particles in any environment.

Microsystem experiments often test the intermediate regime to determine the coherence time.  Figure~1 displays the mean mode 2 population and dipole moment for an atom prepared in mode 2 at time $t=0$ with zero detuning ($\Delta = 0$) and a degree of coupling that produces a reduction rate of $\gamma = 0.25 \Omega_{\text{R}}$. These data are the average of multiple simulated trajectories. Each trajectory was computed by random selection of one of the possible outcomes after each time interval. The asymptotic value is 0.5, equal occupation of both modes, for any finite reduction rate. Note that each test maintains full oscillation amplitude that becomes less synchronized, from test to test, with each reduction event. The average signal of many runs decays exponentially and, confusingly, appears to remain in phase with the ideal oscillation as if each atom decays by some novel mechanism.

In contrast, spontaneous emission from mode 2 to mode 1 may be modeled by two outcomes after each time interval, replacing \eqref{firstincrement} with $\vert \psi(\tau)\rangle = \{(1-\gamma\tau)\chi(\tau), (\gamma\tau)\vert 1 \rangle\}$. This produces a similar average result for low emission rate, but the asymptote drops toward zero, i.e. mode 1 occupation, when the emission rate is much greater than the Rabi frequency.

Other non-ideal dynamics may contribute to mean oscillation decay as well. Such attenuation observed in Refs.~\cite{Kuhr, Jones}, for examples, is attributed to dephasing between individual responses due to varying trap conditions as well as finite temperature. Atom position within the focused laser beam and atom velocity distribution cause Rabi frequency and detuning variation such that oscillation becomes unsynchronized from one test (or qubit) to the next. These effects are reversible in principle so long as the trap is steady. On the other hand, fluctuation in the trapping beams and thermal coupling to the environment cause irreversible dephasing and occasional reduction. Spontaneous emission, reduction and dephasing all produce results of similar character and may be difficult to distinguish. Highly stable trapping beams, wide drive beam relative to the trap, low temperature and weak coupling (i.e. low collision rate indicated by low rate of temperature rise) are all necessary to achieve long coherence time.

\section{Conclusion}

The five anomalous phenomena stated in the introduction are resolved in this dynamic mode picture recognizing spontaneous events as ubiquitous and essential to natural evolution.
Such events disrupt quantum interference and local mode formation, triggering collapse to stable modes.
Interference then may be said to enforce conservation laws, to localize particles in condensed matter, to create the impression of point particles and wave-particle duality, to collapse mode superposition in natural conditions, and to avoid Schr\"{o}dinger's cat paradox. In short, quantum interference creates our reality.

Reduction occurs naturally whenever particles are dense enough to respond quickly to effectively stochastic feedback from their neighbors. We observe only reduced states because measurement involves complex, condensed matter devices in order to produce a macroscopic signal.

All particles reduce rapidly in systems under natural conditions. They are uncorrelated as a result and diffuse among accessible modes because net transition rate between modes is proportional to quantum mechanical rate and the mode occupation. Diffusion from a highly populated group of modes with common macroscopic properties represents friction and all forms of dissipation, transferring bulk motion into heat, while diffusion into such a narrow group is statistically insignificant. Diffusion causes any initial state to evolve toward a steady state, eventually settling, if the environment is undisturbed, into quasi-equilibrium in which each particle species is distributed closely to the limiting case \eqref{EQpopspecies}.
Every system equilibrates at all times through diffusion even when outside activity might drive it away from a steady state.

Thermodynamics is a manifestation of the SMP and inherently quantum mechanical, not an independent theory, as suggested by the names ``classical thermodynamics'' versus ``quantum thermodynamics.'' Classical Thermodynamics and Statistical Mechanics theories were initial attempts to deduce general relations between macroscopic properties with crucial yet still unfounded assumptions made prior to knowledge of microscopic quantum physics. Any theory that relies on results of these original theories to justify their assumptions becomes circular.

Classical properties emerge when the reduction rate greatly exceeds events among individual pairs of particles and the chance of occupying the same mode is small. Any system containing a large enough number of particles in close proximity will appear to exist in one place and not spread spontaneously over time because each particle is localized by its neighbors. Particles at the surface of condensed matter become delocalized when they sublimate into vapor.
In broader terms, classical physics is just a limiting case of quantum mechanics in natural conditions when diffusion of the objects of interest is negligible.

Local conditions determine a particle state and evolution. This region is microscopic in condensed matter except for highly ordered crystalline material phases and near critical points. Distant particles may influence local systems through electromagnetic and gravity fields but do not play a role in local diffusion. This conclusion avoids conundrums in quantum mechanics such as how the state of the universe can be communicated to every particle in real time to maintain the Second Law.

Local fluctuation is crucial for understanding natural behavior. These fluctuations do microscopic work that undermines Lagrangian analysis. Only particles in stable modes behave according to Lagrangian equations of motion for longer than the reduction time. The full panoply of spooky quantum phenomena appear only in carefully controlled environments where reduction is infrequent.

Reduction occurs with sufficient jostling by neighbors, consistent with QDT and ETH theories developed to understand decay in well isolated microsystems by assuming mechanisms with similar stochastic features. The transition from ideal quantum mechanics to thermal dynamics may be understood as the mean effect of reduction events. Both reversible processes and irreversible dissipation may be evident in the intermediate regime when the rates of these mechanisms are comparable.

Many theories proposed to date have hinted at this solution, though founded on assumptions outside of the SMP or not consistently justified. The impasse apparently has been strict adherence to a fixed eigenstate description of the system developed for experiments designed to reveal fundamental physics, when the effect of the environment may be treated as weak perturbation during the experiment. Recognizing that single particle modes must fluctuate provides an endemic mechanism for reduction and thermalization without additional assumptions beyond the SMP.

\end{document}

%% file: Fig_Level2pop.tex
% GNUPLOT: LaTeX picture with Postscript
\begingroup
  \makeatletter
  \providecommand\color[2][]{%
    \GenericError{(gnuplot) \space\space\space\@spaces}{%
      Package color not loaded in conjunction with
      terminal option `colourtext'%
    }{See the gnuplot documentation for explanation.%
    }{Either use 'blacktext' in gnuplot or load the package
      color.sty in LaTeX.}%
    \renewcommand\color[2][]{}%
  }%
  \providecommand\includegraphics[2][]{%
    \GenericError{(gnuplot) \space\space\space\@spaces}{%
      Package graphicx or graphics not loaded%
    }{See the gnuplot documentation for explanation.%
    }{The gnuplot epslatex terminal needs graphicx.sty or graphics.sty.}%
    \renewcommand\includegraphics[2][]{}%
  }%
  \providecommand\rotatebox[2]{#2}%
  \@ifundefined{ifGPcolor}{%
    \newif\ifGPcolor
    \GPcolorfalse
  }{}%
  \@ifundefined{ifGPblacktext}{%
    \newif\ifGPblacktext
    \GPblacktexttrue
  }{}%
  % define a \g@addto@macro without @ in the name:
  \let\gplgaddtomacro\g@addto@macro
  % define empty templates for all commands taking text:
  \gdef\gplbacktext{}%
  \gdef\gplfronttext{}%
  \makeatother
  \ifGPblacktext
    % no textcolor at all
    \def\colorrgb#1{}%
    \def\colorgray#1{}%
  \else
    % gray or color?
    \ifGPcolor
      \def\colorrgb#1{\color[rgb]{#1}}%
      \def\colorgray#1{\color[gray]{#1}}%
      \expandafter\def\csname LTw\endcsname{\color{white}}%
      \expandafter\def\csname LTb\endcsname{\color{black}}%
      \expandafter\def\csname LTa\endcsname{\color{black}}%
      \expandafter\def\csname LT0\endcsname{\color[rgb]{1,0,0}}%
      \expandafter\def\csname LT1\endcsname{\color[rgb]{0,1,0}}%
      \expandafter\def\csname LT2\endcsname{\color[rgb]{0,0,1}}%
      \expandafter\def\csname LT3\endcsname{\color[rgb]{1,0,1}}%
      \expandafter\def\csname LT4\endcsname{\color[rgb]{0,1,1}}%
      \expandafter\def\csname LT5\endcsname{\color[rgb]{1,1,0}}%
      \expandafter\def\csname LT6\endcsname{\color[rgb]{0,0,0}}%
      \expandafter\def\csname LT7\endcsname{\color[rgb]{1,0.3,0}}%
      \expandafter\def\csname LT8\endcsname{\color[rgb]{0.5,0.5,0.5}}%
    \else
      % gray
      \def\colorrgb#1{\color{black}}%
      \def\colorgray#1{\color[gray]{#1}}%
      \expandafter\def\csname LTw\endcsname{\color{white}}%
      \expandafter\def\csname LTb\endcsname{\color{black}}%
      \expandafter\def\csname LTa\endcsname{\color{black}}%
      \expandafter\def\csname LT0\endcsname{\color{black}}%
      \expandafter\def\csname LT1\endcsname{\color{black}}%
      \expandafter\def\csname LT2\endcsname{\color{black}}%
      \expandafter\def\csname LT3\endcsname{\color{black}}%
      \expandafter\def\csname LT4\endcsname{\color{black}}%
      \expandafter\def\csname LT5\endcsname{\color{black}}%
      \expandafter\def\csname LT6\endcsname{\color{black}}%
      \expandafter\def\csname LT7\endcsname{\color{black}}%
      \expandafter\def\csname LT8\endcsname{\color{black}}%
    \fi
  \fi
    \setlength{\unitlength}{0.0500bp}%
    \ifx\gptboxheight\undefined%
      \newlength{\gptboxheight}%
      \newlength{\gptboxwidth}%
      \newsavebox{\gptboxtext}%
    \fi%
    \setlength{\fboxrule}{0.5pt}%
    \setlength{\fboxsep}{1pt}%
\begin{picture}(5384.00,2834.00)%
    \gplgaddtomacro\gplbacktext{%
      \csname LTb\endcsname%
      \put(807,545){\makebox(0,0)[r]{\strut{}\small{0}}}%
      \csname LTb\endcsname%
      \put(807,1557){\makebox(0,0)[r]{\strut{}\small{0.5}}}%
      \csname LTb\endcsname%
      \put(807,2569){\makebox(0,0)[r]{\strut{}\small{1}}}%
      \csname LTb\endcsname%
      \put(905,338){\makebox(0,0){\strut{}\small{0}}}%
      \csname LTb\endcsname%
      \put(1313,338){\makebox(0,0){\strut{}\small{1}}}%
      \csname LTb\endcsname%
      \put(1721,338){\makebox(0,0){\strut{}\small{2}}}%
      \csname LTb\endcsname%
      \put(2130,338){\makebox(0,0){\strut{}\small{3}}}%
      \csname LTb\endcsname%
      \put(2538,338){\makebox(0,0){\strut{}\small{4}}}%
      \csname LTb\endcsname%
      \put(2946,338){\makebox(0,0){\strut{}\small{5}}}%
      \csname LTb\endcsname%
      \put(3354,338){\makebox(0,0){\strut{}\small{6}}}%
      \csname LTb\endcsname%
      \put(3762,338){\makebox(0,0){\strut{}\small{7}}}%
      \csname LTb\endcsname%
      \put(4171,338){\makebox(0,0){\strut{}\small{8}}}%
      \csname LTb\endcsname%
      \put(4579,338){\makebox(0,0){\strut{}\small{9}}}%
      \csname LTb\endcsname%
      \put(4987,338){\makebox(0,0){\strut{}\small{10}}}%
      \put(4579,2367){\makebox(0,0)[l]{\strut{}\small{(a)}}}%
    }%
    \gplgaddtomacro\gplfronttext{%
      \csname LTb\endcsname%
      \put(404,1557){\rotatebox{-270}{\makebox(0,0){\strut{}\small{Mean Mode 2 population}}}}%
      \put(2946,154){\makebox(0,0){\strut{}\small{Time ($2\pi/\Omega_{\text{R}}$)}}}%
    }%
    \gplbacktext
    \put(0,0){\includegraphics{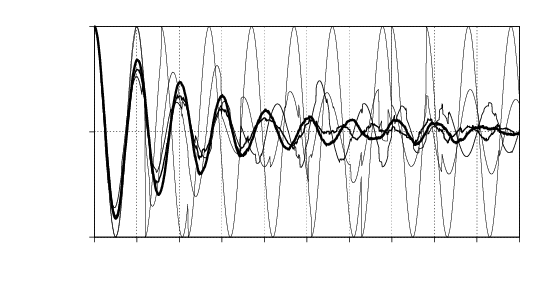}}%
    \gplfronttext
  \end{picture}%
\endgroup

%% file: Fig_Dipole.tex
% GNUPLOT: LaTeX picture with Postscript
\begingroup
  \makeatletter
  \providecommand\color[2][]{%
    \GenericError{(gnuplot) \space\space\space\@spaces}{%
      Package color not loaded in conjunction with
      terminal option `colourtext'%
    }{See the gnuplot documentation for explanation.%
    }{Either use 'blacktext' in gnuplot or load the package
      color.sty in LaTeX.}%
    \renewcommand\color[2][]{}%
  }%
  \providecommand\includegraphics[2][]{%
    \GenericError{(gnuplot) \space\space\space\@spaces}{%
      Package graphicx or graphics not loaded%
    }{See the gnuplot documentation for explanation.%
    }{The gnuplot epslatex terminal needs graphicx.sty or graphics.sty.}%
    \renewcommand\includegraphics[2][]{}%
  }%
  \providecommand\rotatebox[2]{#2}%
  \@ifundefined{ifGPcolor}{%
    \newif\ifGPcolor
    \GPcolorfalse
  }{}%
  \@ifundefined{ifGPblacktext}{%
    \newif\ifGPblacktext
    \GPblacktexttrue
  }{}%
  % define a \g@addto@macro without @ in the name:
  \let\gplgaddtomacro\g@addto@macro
  % define empty templates for all commands taking text:
  \gdef\gplbacktext{}%
  \gdef\gplfronttext{}%
  \makeatother
  \ifGPblacktext
    % no textcolor at all
    \def\colorrgb#1{}%
    \def\colorgray#1{}%
  \else
    % gray or color?
    \ifGPcolor
      \def\colorrgb#1{\color[rgb]{#1}}%
      \def\colorgray#1{\color[gray]{#1}}%
      \expandafter\def\csname LTw\endcsname{\color{white}}%
      \expandafter\def\csname LTb\endcsname{\color{black}}%
      \expandafter\def\csname LTa\endcsname{\color{black}}%
      \expandafter\def\csname LT0\endcsname{\color[rgb]{1,0,0}}%
      \expandafter\def\csname LT1\endcsname{\color[rgb]{0,1,0}}%
      \expandafter\def\csname LT2\endcsname{\color[rgb]{0,0,1}}%
      \expandafter\def\csname LT3\endcsname{\color[rgb]{1,0,1}}%
      \expandafter\def\csname LT4\endcsname{\color[rgb]{0,1,1}}%
      \expandafter\def\csname LT5\endcsname{\color[rgb]{1,1,0}}%
      \expandafter\def\csname LT6\endcsname{\color[rgb]{0,0,0}}%
      \expandafter\def\csname LT7\endcsname{\color[rgb]{1,0.3,0}}%
      \expandafter\def\csname LT8\endcsname{\color[rgb]{0.5,0.5,0.5}}%
    \else
      % gray
      \def\colorrgb#1{\color{black}}%
      \def\colorgray#1{\color[gray]{#1}}%
      \expandafter\def\csname LTw\endcsname{\color{white}}%
      \expandafter\def\csname LTb\endcsname{\color{black}}%
      \expandafter\def\csname LTa\endcsname{\color{black}}%
      \expandafter\def\csname LT0\endcsname{\color{black}}%
      \expandafter\def\csname LT1\endcsname{\color{black}}%
      \expandafter\def\csname LT2\endcsname{\color{black}}%
      \expandafter\def\csname LT3\endcsname{\color{black}}%
      \expandafter\def\csname LT4\endcsname{\color{black}}%
      \expandafter\def\csname LT5\endcsname{\color{black}}%
      \expandafter\def\csname LT6\endcsname{\color{black}}%
      \expandafter\def\csname LT7\endcsname{\color{black}}%
      \expandafter\def\csname LT8\endcsname{\color{black}}%
    \fi
  \fi
    \setlength{\unitlength}{0.0500bp}%
    \ifx\gptboxheight\undefined%
      \newlength{\gptboxheight}%
      \newlength{\gptboxwidth}%
      \newsavebox{\gptboxtext}%
    \fi%
    \setlength{\fboxrule}{0.5pt}%
    \setlength{\fboxsep}{1pt}%
\begin{picture}(5384.00,2834.00)%
    \gplgaddtomacro\gplbacktext{%
      \csname LTb\endcsname%
      \put(807,589){\makebox(0,0)[r]{\strut{}\small{0}}}%
      \put(807,985){\makebox(0,0)[r]{\strut{}\small{0.1}}}%
      \put(807,1381){\makebox(0,0)[r]{\strut{}\small{0.2}}}%
      \put(807,1777){\makebox(0,0)[r]{\strut{}\small{0.3}}}%
      \put(807,2173){\makebox(0,0)[r]{\strut{}\small{0.4}}}%
      \put(807,2569){\makebox(0,0)[r]{\strut{}\small{0.5}}}%
      \put(905,381){\makebox(0,0){\strut{}\small{0}}}%
      \put(1313,381){\makebox(0,0){\strut{}\small{1}}}%
      \put(1721,381){\makebox(0,0){\strut{}\small{2}}}%
      \put(2130,381){\makebox(0,0){\strut{}\small{3}}}%
      \put(2538,381){\makebox(0,0){\strut{}\small{4}}}%
      \put(2946,381){\makebox(0,0){\strut{}\small{5}}}%
      \put(3354,381){\makebox(0,0){\strut{}\small{6}}}%
      \put(3762,381){\makebox(0,0){\strut{}\small{7}}}%
      \put(4171,381){\makebox(0,0){\strut{}\small{8}}}%
      \put(4579,381){\makebox(0,0){\strut{}\small{9}}}%
      \put(4987,381){\makebox(0,0){\strut{}\small{10}}}%
      \put(4579,2371){\makebox(0,0)[l]{\strut{}\small{(b)}}}%
    }%
    \gplgaddtomacro\gplfronttext{%
      \csname LTb\endcsname%
      \put(404,1579){\rotatebox{-270}{\makebox(0,0){\strut{}\small{Mean dipole magnitude}}}}%
      \put(2946,154){\makebox(0,0){\strut{}\small{Time ($2\pi/\Omega_{\text{R}}$)}}}%
    }%
    \gplbacktext
    \put(0,0){\includegraphics{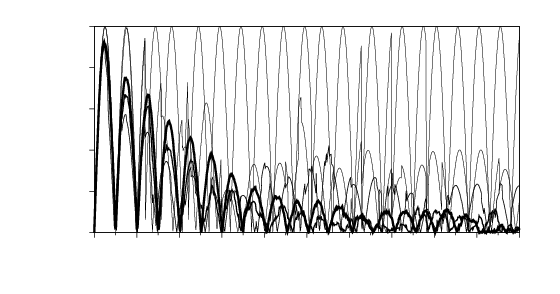}}%
    \gplfronttext
  \end{picture}%
\endgroup

%% file: fromQDTtoTD_Brownell2024.bbl
\begin{thebibliography}{99}

\bibitem{Poincare}{ Poincar\'{e}, H. On the three-body problem and the equations of dynamics. Acta Math. 13.1 (1890). }

\bibitem{Sechoes}{ Hahn, E. L. Spin echoes. Physical review 80.4 (1950): 580. }

\bibitem{Pechoes}{ Abella, I. D., N. A. Kurnit, and S. R. Hartmann. Photon echoes. Physical review 141.1 (1966): 391. }

\bibitem{Goldstein}{ Goldstein, H. \textit{Classical Mechanics} (2nd edition), Addison Wesley, 1980. }

\bibitem{Uffink}{ Uffink, J. Compendium of the foundations of classical statistical physics. J. Butterfield and J. Earman (eds), Philosophy of Physics, Amsterdam: North Holland (2007): 923-1047. }

\bibitem{Boltzmann}{ Boltzmann, L. Wissenschaftliche Abhandlungen Vol. II, F. Hasen\"{o}hrl (ed.) Leipzig (1909): 165-6. Reissued New York: Chelsea, 1969. See Ref.~\cite{Uffink}, p. 56. }

\bibitem{Gibbs}{ Gibbs, J. W. \textit{Elementary principles in statistical mechanics: developed with especial reference to the rational foundations of thermodynamics}, C. Scribner's sons, 1902.}

\bibitem{Einstein}{ Einstein, A. Theory of opalescence of homogeneous liquids and mixtures of liquids in the vicinity of the critical state. Ann. Physik 33 (1910): 1275. }

\bibitem{Jaynes}{ Jaynes, E. T. Information theory and statistical mechanics. Physical review 106.4 (1957): 620. }

\bibitem{MWI}{ Vaidman, L. Many-Worlds Interpretation of Quantum Mechanics, The Stanford Encyclopedia of Philosophy (Fall 2021 Edition), Edward N. Zalta (ed.). https://plato.stanford.edu/archives/fall2021/entries/qm-manyworlds/. }

\bibitem{Zurek}{ Zurek, W. H. Quantum Theory of the Classical: Einselection, Envariance, Quantum Darwinism and Extantons. Entropy 24.11 (2022): 1520. }

\bibitem{Srednicki1994}{ Srednicki, M. Chaos and quantum thermalization. Physical review E 50.2 (1994): 888-901. }

\bibitem{Alessio}{ D'Alessio, L., Y. Kafri, A. Polkovnikov, and M. Rigol. From quantum chaos and eigenstate thermalization to statistical mechanics and thermodynamics. Advances in Physics 65.3 (2016): 239-362. }

\bibitem{Ashida}{ Ashida, Y., Z. Gong, and M. Ueda. Non-hermitian physics. Advances in Physics 69.3 (2020): 249-435. }

\bibitem{Bassi}{Bassi, A., and G.C. Ghirardi. Dynamical reduction models. Physics Reports 379.5-6 (2003): 257-426.}

\bibitem{Brownell:transport}{ Brownell, J. H. Thermodynamic transport equations derived from first principles. doi.org/10.5281/zenodo.6815735 (2022). }

\bibitem{experiments}{ Millen, J., and B. A. Stickler. Quantum experiments with microscale particles. Contemporary Physics 61.3 (2020): 155-168.}

\bibitem{Isar}{ Isar, A., A. Sandulescu, H. Scutaru, E. Stefanescu, and W. Scheid. Open quantum systems. International Journal of Modern Physics E, 3.2 (1994), 635-714. }

\bibitem{Vega}{ De Vega, I., and D. Alonso. Dynamics of non-Markovian open quantum systems. Reviews of Modern Physics, 89.1 (2017): 015001. }

\bibitem{Ciccarello}{ Ciccarello, F., S. Lorenzo, V. Giovannetti, and G. M. Palma. Quantum collision models: Open system dynamics from repeated interactions. Physics Reports, 954 (2022): 1-70. }

\bibitem{Schlosshauer}{Schlosshauer, M. Quantum decoherence. Physics Reports 831 (2019): 1-57. }

\bibitem{Guhr}{ Guhr, T., A. Müller–Groeling, and H. A. Weidenmüller. Random-matrix theories in quantum physics: common concepts. Physics Reports 299, no. 4-6 (1998): 189-425. }

\bibitem{Berry}{ Berry, M.V. Regular and irregular semiclassical wavefunctions. J. Phys. A 10 (1977): 2083. }

\bibitem{Anderson}{ Anderson, P. W. Absence of Diffusion in Certain Random Lattices. Physical Review 109.5 (1958): 1492–1505. }

\bibitem{Plenio}{Plenio, M. B., and P. L. Knight. The quantum-jump approach to dissipative dynamics in quantum optics. Reviews of Modern Physics 70.1 (1998): 101.}

\bibitem{Shulman}{ Schulman, L. S. Jump time and passage time: The duration of a quantum transition. In Time in Quantum Mechanics, pp. 107-128. Berlin, Heidelberg: Springer Berlin Heidelberg, 2007. }

\bibitem{delaPena}{ de la Peña, L., A. M. Cetto, and A. Valdés-Hernández. How fast is a quantum jump? Physics Letters A 384.34 (2020): 126880. }

\bibitem{QC}{ Ayral, T., P. Besserve, D. Lacroix, and E. A. R. Guzman. Quantum computing with and for many-body physics. The European Physical Journal A 59, no. 10 (2023): 227. }

\bibitem{Frigg2021a}{ Frigg, R., and C. Werndl. Equilibrium in Boltzmannian statistical mechanics. The Routledge Companion to Philosophy of Physics. Routledge (2021): 403-413. }

\bibitem{Helbig}{ Helbig, T., T. Hofmann, R. Thomale, and M. Greiter. Theory of Eigenstate Thermalisation.
arXiv:2406.01448 [quant-ph] (2024). }

\bibitem{Moudgalya}{ Moudgalya, S., B. A. Bernevig, and N. Regnault. Quantum many-body scars and Hilbert space fragmentation: a review of exact results. Reports on Progress in Physics 85.8 (2022): 086501. }

\bibitem{Chandran}{ Chandran, A., T. Iadecola, V. Khemani, and R. Moessner. Quantum many-body scars: A quasiparticle perspective. Annual Review of Condensed Matter Physics 14 (2023): 443-469. }

\bibitem{Kuhr}{ Kuhr, S., W. Alt, D. Schrader, I. Dotsenko, Y. Miroshnychenko, A. Rauschenbeutel, and D. Meschede. Analysis of dephasing mechanisms in a standing-wave dipole trap. Physical Review A, 72.2 (2005): 023406. }

\bibitem{Jones}{ Jones, M. P., J. Beugnon, A. Gaëtan, J. Zhang, G. Messin, A. Browaeys, and P. Grangier. Fast quantum state control of a single trapped neutral atom. Physical Review A, 75.4 (2007): 040301. }

\bibitem{Deutsch}{ Deutsch, J. M. Eigenstate thermalization hypothesis. Reports on Progress in Physics 81.8 (2018): 082001. }

\bibitem{Zeno}{ Nakanishi, T., K. Yamane, and M. Kitano. Absorption-free optical control of spin systems: the quantum Zeno effect in optical pumping. Physical Review A. 65.1 (2001): 013404. }

\bibitem{DF}{ Darwin, C. G., and R. H. Fowler. ``XLIV. On the partition of energy.'' The London, Edinburgh, and Dublin Philosophical Magazine and Journal of Science, 44.261 (1922): 450–479. https://doi.org/10.1080/14786440908565189 }


\end{thebibliography}
